\documentclass{article}
\usepackage[utf8]{inputenc}
\usepackage{amsmath}
\usepackage{amsfonts}
\usepackage{amssymb}
\usepackage{mathtools}
\DeclarePairedDelimiter{\ceil}{\lceil}{\rceil}
\DeclarePairedDelimiter{\floor}{\lfloor}{\rfloor}
\usepackage{xcolor,colortbl}
\usepackage{adjustbox}

\title{It Is Easy For Multi-Issue Bundles To Advance Anti-Democratic Agendas}
\author{Matthew I. Jones, Matthew Chervenak, Nicholas A. Christakis}
\date{}

\begin{document}

\maketitle

\section*{Abstract}

When confronted with a host of issues, groups often save time and energy by compiling many issues into a single bundle when making decisions. This reduces the time and cost of group decision-making, but it also leads to suboptimal outcomes as individuals lose the ability to express their preferences on each specific issue. We examine this trade-off by quantifying the value of bundled voting compared to a more tedious issue-by-issue voting process. 
Our research investigates multi-issue bundles and their division into multiple subbundles, confirming that bundling generally yields positive outcomes for the group. However, bundling and issue-by-issue voting can easily yield opposite results regardless of the number of votes the bundle receives.
Furthermore, we show that most combinations of voters and issues are vulnerable to manipulation if the subundling is controlled by a bad actor. By carefully crafting bundles, such an antagonist can achieve the minority preference on almost every issue. Thus, naturally occurring undemocratic outcomes may be rare, but they can be easily manufactured.
To thoroughly investigate this problem, we employ three techniques throughout the paper: mathematical analysis, computer simulations, and the analysis of American voter survey data.
This study provides valuable insights into the dynamics of bundled voting and its implications for group decision-making. By highlighting the potential for manipulation and suboptimal outcomes, our findings add another layer to our understanding of voting paradoxes and offer insights for those designing group decision-making systems that are safer and more fair.

\section{Introduction}
Groups are often required to make decisions about multiple issues at the same time. In a world in which voters have unlimited time and energy, every issue could be decided in isolation, thus ensuring that the will of the majority is carried out on every issue and maximizing value for the group as a whole. However, for a host of reasons ranging from the intrinsic relatedness of some issues to time, convenience, and self-interested manipulation of the system, groups often make decisions on a range of issues with one vote. This requires placing multiple issues into a single \emph{bundle} and giving each group member one vote for the entire set of issues.

Our recurring toy example will be a group of friends ordering a single pizza with many available toppings. To decide which toppings to order, the group could go down the list of toppings and do a vote on each. ``Pepperoni, yes. Mushrooms, no. Sausage, no...'' and so on. This process would take a long time, but in the end, the group should be satisfied with the pizza, as each topping was added (or left off) by the will of the majority. 

Now suppose everybody is hungry and rushed. Instead of taking the time to vote on each topping separately, one person might propose a vote between ordering the pizza with everything on it or nothing on it. If most members of the group prefer the ``supreme'' pizza to a plain cheese, the group orders a pizza with all the toppings, even if some are actually opposed by the majority. Is this pizza a good choice for the group? How does this pizza compare to the best possible pizza the group could have ordered? Was the sacrifice in quality worth the speed and ease with which the group was able to make this choice? Here, we provide a mathematical framework to quantify how group utility is impacted by bundling multiple issues together and explore other more serious applications.

Multi-issue bundling carries the risk of manipulation. A bad actor with control of the voting agenda of the group can choose to place certain issues in a bundle to prevent popular issues from passing. Therefore, throughout this analysis, we will consider a theoretical non-voting antagonistic agenda-setter (NVAAS) who has a preference for the unpopular position on every issue. Of course, the agenda-setter could instead prefer the popular position on every issue. This would end very well for the group, but this is unlikely to happen and impossible to guarantee, so we focus on the worst-case scenario in which the agenda-setter is opposed to the group on every issue. This NVAAS will attempt to bundle issues together so that the final voting outcome is as far from the issue-by-issue majority rule as possible. Our NVAAS could be a disgruntled pizza shop owner who wants to sell generally unappetizing toppings by bundling them together, or, in more realistic settings, a lawmaker trying to pass unpopular legislation or a carmaker attempting to sell unwanted amenities by bundling them with popular ones.

Early investigations into bundling demonstrated that voting on bundles can benefit the minority, rather than majority, position on each issue \cite{hillinger_1971}, but that if division of the question (dividing a bundle into its constituent parts) is automatically applied to a bundle, the bundle made up of the median position on each issue is the Condorcet winner \cite{kadane_1972}. Later, a series of related ``voting paradoxes'' 
expanded upon the distinctions between bundle and single-issue votes, including the Ostrogorski Paradox\cite{RAE_DAUDT_1976}, Anscombe's Paradox \cite{Wagner_1983}, and the Paradox of Multiple Elections \cite{Brams_Kilgour_Zwicker_1998}. However, these paradoxes are sometimes dismissed as being of little concern, either because they are trivial or extremely unlikely \cite{Gehrlein_Merlin_2020}. Our work runs counter to this narrative, showing that anti-democratic outcomes are quite flexible and can be easily manufactured and maintained by an NVAAS.

Unsurprisingly, political scientists have studied the bundling of multiple issues in the context of legislation, an area full of complex bills and high-stakes decisions with the potential for manipulation \cite{Riker_1986}. There is substantial empirical literature studying the omnibus bill in particular \cite{krutz_2001, krutz_2021}, but examining the multifaceted nature of bills is of general interest to the political science community \cite{rundquist_strom_1987}. The compilation of multiple legislative issues into a single bundle has also been proposed as a work-around to avoid the detrimental effect of loss aversion \cite{milkman_mazza_shu_tsay_bazerman_2012}. Bundling can also (for better or worse) save legislators from having to read and consider every minor detail of a bill \cite{volokh_2011}, and it can simplify decision-making for single issue voters \cite{congleton_1991}. Other recent work considers how multiple issues can be formed into bundles by strategic actors seeking to maximize their own utility \cite{alonso_camara_2016, camara_eguia_2017}. 

Of course, there are other fields besides politics in which many issues are grouped into a single choice. The bundling of amenities, like parking included in rent, results in a non-optimal distribution of resources because renters without cars are forced to pay for parking spaces they don't need \cite{gabbe_pierce_2016}. The same holds for the packages of amenities, called trim levels, that are sold with new cars \cite{doyle_2018}. While sometimes counter-productive, bundling products is a good way to increase sales of individual products. Consumers may be willing to buy a whole collection of products to get the one item they really desire \cite{priessner_hampl_2020}, but consumers purchasing a bundle of products expect them to be discounted because they are in a bundle \cite{heeler_nguyen_buff_2007}. While companies claim these bundles are for the benefit of consumers, it is worth considering them as NVAASes, aiming to maximize profit at the consumer's expense.

Here, we focus on the intrinsic value that bundling creates for the whole group when individuals are voting sincerely, with or without an NVAAS. We show that, in the average case, large bundles have a small but positive value, suggesting there is some merit to placing large number of issues together. However, when the bundling is controlled by a bad-faith actor, it is a powerful tool to subvert the majority will on most issues. Using techniques from the study of gerrymandering, we confirm that bundling can lead to anti-democratic outcomes when antagonistic agents are given complete information and control of the bundling agenda. Finally, we use American voter survey data and demonstrate how this technique could be used to manipulate real groups of real people making decisions on real political issues. Our results have consequences for group decision-making, particularly in the area of politics.

\section{Methods}
In this section, we describe our model of multi-issue voting and highlight the main features of the model. We begin with $n$ group members or voters and $m$ binary (we call the two choices ``yes'' and ``no'' for simplicity) issues. To avoid some minor idiosyncrasies from ties and tiebreaking, we assume that $n$ and $m$ are both odd unless specifically mentioned. 

We make two assumptions in our model: \emph{separability} and equal utility. Separability means these issues are independent for each voter, so a voter's preferences on one issue do not depend on the outcome of another issue. This may be true for some issues (a car buyer may only be interested in paying for leather seats if they get other luxury amenities like heated seats and large infotainment screens), but not for others (the quantity of cup holders is unlikely to affect a buyer's desire for advanced safety features). Equal utility means that all issues have the same value to every voter which is normalized to 1 for simplicity. These are standard assumptions when studying multi-issue voting, and in the Discussion section, we consider these assumptions at greater depth, why they are necessary, and what they imply about the model.

\subsection{Bundled Vote Scores}

Table \ref{tab:profile1} shows an example of a utility profile with five voters and three issues. Voters $v_1$, $v_2$, and $v_3$ have net positive utility from the bundle and therefore all three issues pass, even though issues $t_2$ and $t_3$ are opposed by a majority of voters. Was this a good outcome for the group, or would it have been better to vote down the whole bundle? We need a measure to determine how much value is generated for the group by a decision to accept or reject a bundle of issues, so we introduce two metrics: utility scores and issue scores.

\begin{table}
    \centering
    \begin{tabular}{r||c|c|c|}
         &  $t_1$ & $t_2$ & $t_3$\\
         \hline \hline
        $v_1$ & 1 & 1 & 1\\
        \hline
        $v_2$ & 1 & $\cellcolor{red!25}-1$ & 1 \\
        \hline
        $v_3$ & 1 & 1 & $\cellcolor{red!25}-1$ \\
        \hline
        $v_4$ & 1 & $\cellcolor{red!25}-1$ & $\cellcolor{red!25}-1$ \\
        \hline
        $v_5$ & $\cellcolor{red!25}-1$ & $\cellcolor{red!25}-1$ & $\cellcolor{red!25}-1$ \\
        \hline
    \end{tabular}
    \caption{A utility profile for five voters and three issues. Three voters ($v_1$, $v_2$, and $v_3$) support the entire bundle so it will pass, but only one of the issues, $t_1$, is supported by a majority of voters. Throughout this paper, the cells with negative utility are shaded red to increase readability.}
    \label{tab:profile1}
\end{table}

The utility score begins with a single basic idea: the utility for a group is just the sum of utilities of each of the group's members. Therefore, if a bundle is passed, the utility for the group is the sum of all of the entries in the utility profile. If the bundle fails, it is the negative of this sum. Finally, we normalize this value by dividing by the total number of voter-issue pairs, $nm$, so that this measure takes values between $-1$ and $1$. The utility profile in Table \ref{tab:profile1} has a utility score of $\frac{8-7}{3\cdot 5} = \frac{1}{15}$. 

On the other hand, proponents of democracy could argue that the goal of group decision-making is not to maximize social utility but to carry out the will of the majority on every issue. Instead of looking to maximize utility, a good group decision process aims to satisfy the majority on as many \emph{issues} as possible. The issue score is the number of issues where the bundled choice agrees with issue-by-issue voting, subtracted by the number of issues where the two disagree, and normalized by dividing by the number of issues. In Table \ref{tab:profile1}, the bundle passes, but only one of the three issues would pass by itself. Therefore, the bundle's issue score is $\frac{1-2}{3} = - \frac{1}{3}$.

Though these scores are clearly similar, they are not always in agreement. The profile in Table \ref{tab:profile1} has a positive utility score, suggesting that approving the bundle was the correct decision. But it also has a negative issue score, suggesting the bundle should have been rejected. This is one of many profiles where these two measures lead to opposite conclusions, so we will consider both.

\subsection{Voter Utility Models}

These scores allow us to study bundled voting, but we still need a method of forming utility profiles like Table \ref{tab:profile1} to compute the scores. We focus on two models of voter utility formation. 

The first is the independent, identically distributed (IID) model \cite{xefteris_ziros_2017, herrera_morelli_palfrey_2014}. In this model, the value of each cell in the utility profile is $+1$ with probability $p$ and $-1$ with probability $1-p$. This model is analytically tractable, and all the closed-form equations that follow use the IID model. In the special case of $p = \frac{1}{2}$, each utility is a coin flip, and the entire bundle will be very competitive. Approximately half of the issues will be supported and half opposed by the majority even in the limit of large $n$ and large $m$. Later, we show that this highly competitive domain produces the lowest-value bundles.

For a higher-fidelity (and less tractable) model, we use the well-known spatial model of voting \cite{jones_sirianni_fu_2022, poole_rosenthal_1997}. For each issue, we randomly choose two points in a $k$-dimensional hypercube to be the ``yes'' and ``no'' positions. Each voter is also assigned a random ideal point, and her utility for a given issue is determined by the closeness of the voter to the ``yes'' and ``no'' points for that issue. Pizza toppings could be placed in a three-dimensional space by rating them on how salty, sweet, and savory they are. A diner who likes salty and savory things and dislikes foods that are too sweet may prefer toppings like bacon and pepperoni, while another diner who prefers sweet over salty may vote down bundles with things like olives, instead supporting toppings like pineapple or peppers. In this model, utility profiles across voters are highly correlated since two voters with similar points in space will agree on almost all issues. The results shown use $k=2$ dimensions, but the results do not significantly change when using higher dimensions (see Supplementary Information for more details).

To model a growing majority in the IID model, we simply adjusted the parameter $p$. As $p$ grew larger, more utilities in the profile are positive, suggesting the voters are becoming more similar and the issues more popular. For the spatial model, we introduce a parameter $q$ that controls the distribution of ideal points in space. When $q=\frac{1}{2}$, all points are uniformly distributed across the entire space, and when $q$ grows close to $1$, almost all voters will have positive utility on almost all issues. More details on the spatial model and the parameter $q$ are in the Supplementary Information.

\section{Results}

We divide our results into three sections. First, we examine the implications of placing all issues in a single bundle. Then, we consider splitting a bundle into two \emph{subbundles} of popular and unpopular issues, and finally, we generalize this by considering any number of subbundles. In each case, we study the average effect of the bundling process and determine the potential for manipulation by well-informed antagonistic agents. At the end, we test these methods on real data and confirm that these results hold in realistic settings. Our results incorporate large ensembles of computer-generated profiles for simulations alongside mathematical analyses.

\subsection{Single Bundle}

Placing a larger number of issues into a single bundle provides two significant advantages for the decision-maker: time and processing requirements. Returning to our pizza example, no group wants to spend precious minutes voting on each individual topping when that time could be spent eating instead. Furthermore, bundling can simplify the decision process: a voter who loves veggies and sees a proposed pizza with broccoli, spinach, and peppers may not even need to think about the rest of the toppings because she already knows that she will enjoy the pizza. These are tangible benefits, but bundling also has a significant downside. With a bundle, voters are no longer able to provide any information about which issues they support and which they oppose. As a result, voters are forced to support issues with negative utility and vice versa. Depending on the situation, this can be a serious impediment to social value.

In our hypothetical group trying to order a pizza, if the group is evenly divided on each topping, then about half the group will be unhappy with every decision. However, suppose the group is entirely made up of vegetarians. In this case, the group will unanimously reject bacon as a topping, pleasing everyone. They will do the same with all of the meat toppings, and accept many or all of the veggie toppings. As the issues become less divisive and most issue-by-issue votes turn out the same way, a bundled vote will align with a higher fraction of the issue-by-issue votes. 

Similarly, in the IID model, when $n$ and $m$ are large and $p \neq \frac{1}{2}$, each issue vote will go the same way. Each issue has roughly $pn$ supporters and $(1-p)n$ detractors. The expected utility of passage is 

\begin{equation}\label{eq:analytic_unbalanced}
    E(\text{Utility Score}) = m | \frac{pn - (1-p)n}{nm}| = |2p-1|
\end{equation}

With both models and both measures of bundle value, we see an improvement in score as the majority becomes stronger. We conclude that a bundled vote has positive expected value and that the value increases as groups become less heterogeneous. As one faction gains influence and agenda-setting control, bundling erases less value. For the remainder of the paper, we will consider only the highly competitive (and more interesting) case where $p=\frac{1}{2}$ or $q=\frac{1}{2}$.

\begin{figure}
    \centering
    \includegraphics[width = \textwidth]{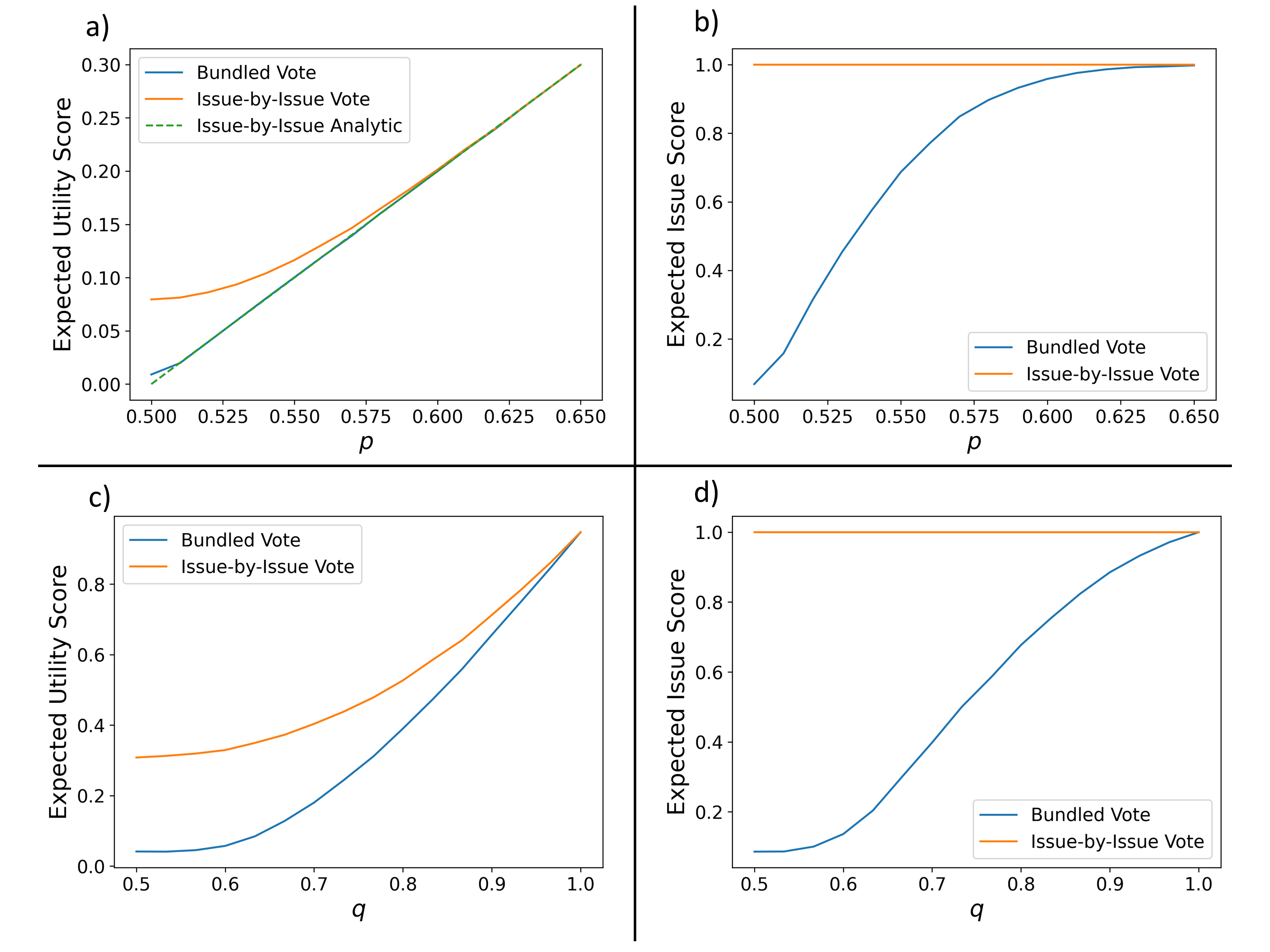}
    \caption{The effect of a strong majority in the IID model (a,b) and the 2D spatial model (c,d) on utility scores (a,c) and issue scores (b,d). The IID model was adjusted by varying $p$ between $\frac{1}{2}$ and $\frac{13}{20}$. The spatial model moved $q$ from $\frac{1}{2}$ to $1$. In (a), simulations confirm our analytic results in Equation \eqref{eq:analytic_unbalanced}, except at $p=\frac{1}{2}$, where Equation \eqref{eq:analytic_score} holds. Parameter values: $n = 101, m=51$.}
    \label{fig:shifting_majority}
\end{figure}

In the Supplementary Information, we derive the expected utility score for a bundle with a large number of voters and issues under the IID model with $p=\frac{1}{2}$ to be

\begin{equation} \label{eq:analytic_score}
    E(\text{utility score})= \frac{2}{\pi\sqrt{nm}}.
\end{equation}

\begin{figure}
    \centering
    \includegraphics[width=\textwidth]{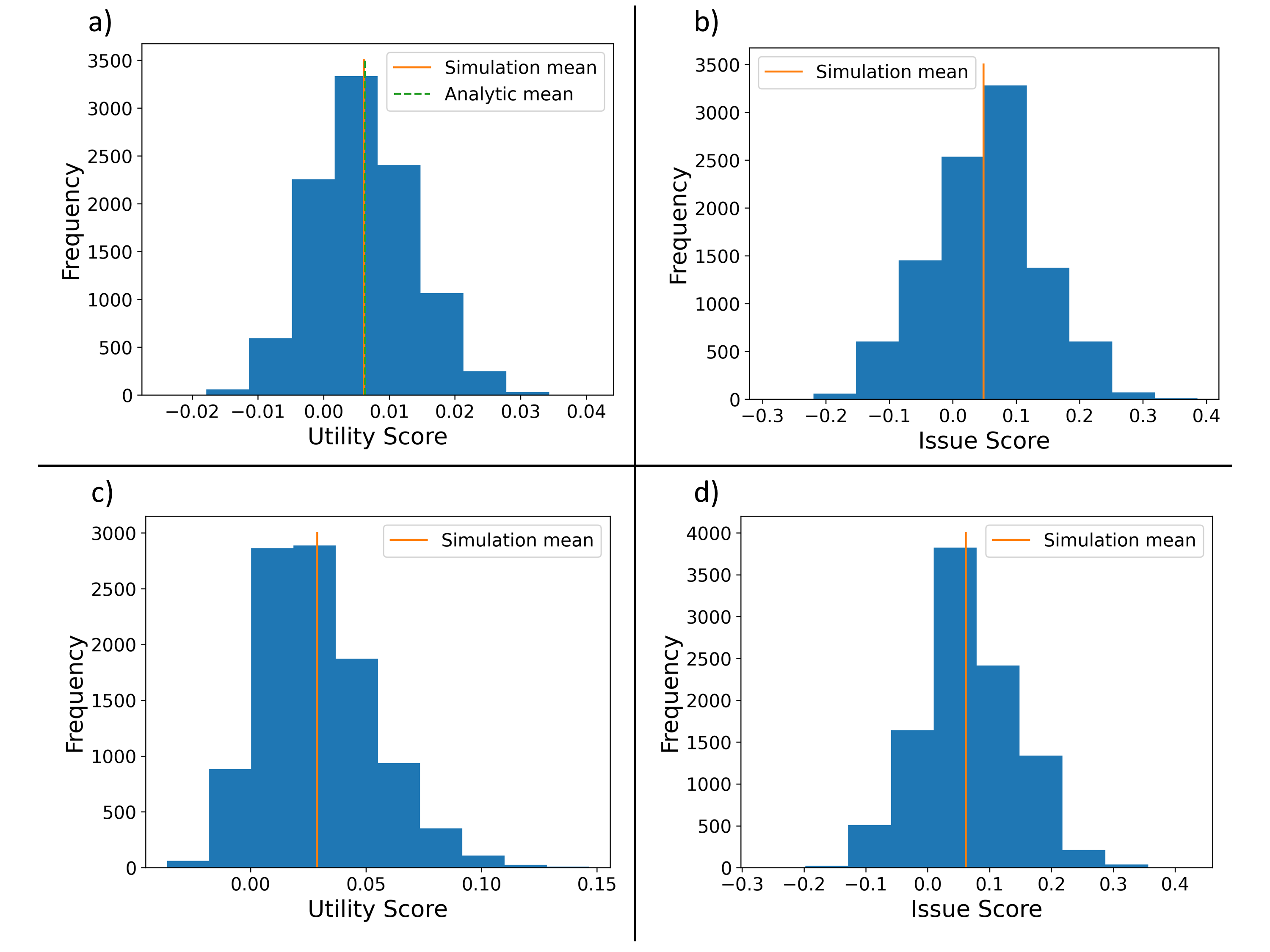}
    \caption{Distribution of utility scores (a,c) and issues scores (b,d) for the $p=\frac{1}{2}$ IID (a,b) and $p = \frac{1}{2}$ 2D spatial (c,d) models of voter utility. Each was generated with 10,000 simulated bundles where $n = m = 101$. Notice the excellent agreement on mean utility score for the IID model in (a) between the analytic approximation from  Equation \eqref{eq:analytic_score} and the simulations.}
    \label{fig:average_scores}
\end{figure}

This slight positive value of bundled votes occurs in perfectly balanced competition, where the expected utility of a bundle is exactly zero. It may seem that these two facts are in contradiction, but they are not. If all the pizza toppings are bad, the group gets a cheese pizza (positive value for not having unpopular toppings) and if the toppings are all good, the group gets the supreme pizza with all the delicious toppings (also positive value). Either way, the utility score will usually be positive. See the Supplementary Information for more details about the value of this marginal utility.

With simulations, we see the distribution of both models and both scores in Figure \ref{fig:average_scores}. In all panels, the average is small but positive, suggesting that bundles are at least slightly effective, and that this conclusion is robust to model parameters.

All the results so far use random bundles. Instead, suppose an NVAAS, perhaps a pizza maker trying to sell unpopular ingredients, gets to custom design a bundle similar to specialty pizzas like meat lovers, chicken-bacon-ranch, etc. If these ingredients are all opposed by a majority of the group, how much can the malevolent agent accomplish with the limited apparatus of bundling? As an example, consider the utility profile in Table \ref{tab:profile2}; the entire bundle is voted down unanimously, but only two of the issues are actually opposed by a majority. Using a single bundle, our NVAAS successfully prevented seven popular issues from passing! This example is similar to the Ostrogorski paradox, in which the winning party has minority views on every issue \cite{RAE_DAUDT_1976}.

\begin{table}
    \centering
    \small
    \begin{tabular}{r||c|c|c|c|c|c|c|c|c|}
         & $t_1$ & $t_2$ & $t_3$ & $t_4$ & $t_5$ & $t_6$ & $t_7$ & $t_8$ & $t_9$
         \\
         \hline \hline
         $v_1$ & $\cellcolor{red!25}-1$ & $\cellcolor{red!25}-1$ & $1$ & $1$ & $1$ & $1$ & $\cellcolor{red!25}-1$ & $\cellcolor{red!25}-1$ & $\cellcolor{red!25}-1$ \\
         \hline
         $v_2$ & $\cellcolor{red!25}-1$ & $\cellcolor{red!25}-1$ & $\cellcolor{red!25}-1$ & $1$ & $1$ & $1$ & $1$ & $\cellcolor{red!25}-1$ & $\cellcolor{red!25}-1$ \\
         \hline
         $v_3$ & $\cellcolor{red!25}-1$ & $\cellcolor{red!25}-1$ & $\cellcolor{red!25}-1$ & $\cellcolor{red!25}-1$ & $1$ & $1$ & $1$ & $1$ & $\cellcolor{red!25}-1$ \\
         \hline
         $v_4$ & $\cellcolor{red!25}-1$ & $\cellcolor{red!25}-1$ & $\cellcolor{red!25}-1$ & $\cellcolor{red!25}-1$ & $\cellcolor{red!25}-1$ & $1$ & $1$ & $1$ & $1$ \\
         \hline
         $v_5$ & $\cellcolor{red!25}-1$ & $\cellcolor{red!25}-1$ & $1$ & $\cellcolor{red!25}-1$ & $\cellcolor{red!25}-1$ & $\cellcolor{red!25}-1$ & $1$ & $1$ & $1$ \\
         \hline
         $v_6$ & $\cellcolor{red!25}-1$ & $\cellcolor{red!25}-1$ & $1$ & $1$ & $\cellcolor{red!25}-1$ & $\cellcolor{red!25}-1$ & $\cellcolor{red!25}-1$ & $1$ & $1$ \\
         \hline
         $v_7$ & $\cellcolor{red!25}-1$ & $\cellcolor{red!25}-1$ & $1$ & $1$ & $1$ & $\cellcolor{red!25}-1$ & $\cellcolor{red!25}-1$ & $\cellcolor{red!25}-1$ & $1$ \\
         \hline
    \end{tabular}
    \caption{A heavily manipulated utility profile of seven voters and nine issues. Negative utilities are highlighted to increase readability. Despite unanimous opposition to the bundle, only two of the nine issues are not majority-supported.}
    \label{tab:profile2}
\end{table}

The utility score for Table \ref{tab:profile2} is a respectable $\frac{1}{9}$, while the issue score is $-\frac{5}{9}$. The reason for this discrepancy in score is clear from the pattern of positive and negative utilities in the utility profile. There is enough positive utility for seven of the issues to have a slim majority support, but they are spread out across all seven voters evenly, so each voter narrowly votes down the bundle. Table \ref{tab:profile2} was created using the same ``packing'' and ``cracking'' techniques that are used to create gerrymandered political redistricting plans \cite{deford_eubank_rodden_2021}.

Because these utility profiles can be strongly gerrymandered in this way, the fraction of voters supporting a bundle provides very limited information about how few or how many of the issues have majority support. Suppose we have a bundle with $n$ voters and $m$ issues, and $\overline{n}$ of the voters support the bill. When $\overline{n}$ is small,

\begin{equation}\label{eq:bound1} \text{Max number of issues supported:  } \floor*{
\frac{(n-\overline{n})(m-1)}{n+1-2\overline{n}}
}.\end{equation}

On the other hand, when $\overline{n}$ is close to $n$,
\begin{equation}\label{eq:bound2} \text{Min number of issues supported:  } \ceil*{
\frac{\overline{n}(m+1) - m (n-1)}{n+1}
}.\end{equation}

See the Supplementary Information for the derivation. Critically, these bounds are extremely weak. For most values of $\overline n$, these bounds become zero and $n$, meaning we can say nothing about the number of majority-supported issues for certain. For example, if a bundle of 15 issues is supported by only 3 of 21 voters, it is possible that all issues could be majority-supported. Even with unanimous support or dissent, we can only be sure that two or three of the issues are actually decided correctly by the bundled vote. Collecting multiple issues into a single bundle could save a group time and energy, but it also exposes them to potentially severe manipulation by a well-informed bad actor, and not even unanimous agreement can provide a guarantee in the value of the decision. These bounds are related to, but distinct from, previous work categorizing the conditions under which voting paradoxes can occur \cite{Deb_Kelsey_1987, Wagner_1984}.

\subsection{Two Subbundles}

Next, we consider the implications of breaking a bundle into two or more subbundles. It is tempting to think that bundle division could be a magic bullet. If we can somehow quickly detect which issues should be passed and which should be rejected, we can put them into two separate subbundles, pass the majority-supported subbundle, and reject the minority-supported subbundle to achieve a perfect issue score of 1. Unfortunately, even if we ignore the question of how to determine which issues have majority support and which do not, an NVAAS could manipulate both in a similar manner to Table \ref{tab:profile2}. 

\begin{table}[]
    \centering
    \begin{tabular}{m{0.25cm}||m{0.2cm}|m{0.2cm}|m{0.2cm}|m{0.2cm}|m{0.2cm}|m{0.2cm}|m{0.2cm}|m{0.2cm}|m{0.2cm}|m{0.2cm}|m{0.2cm}||m{0.2cm}|m{0.2cm}|m{0.2cm}|m{0.2cm}|m{0.2cm}|m{0.2cm}|m{0.2cm}|m{0.2cm}|m{0.2cm}|m{0.2cm}|}
         & $t_1$ & $t_2$ & $t_3$ & $t_4$ & $t_5$ & $t_6$ & $t_7$ & $t_8$ & $t_9$ & $t_{10}$ & $t_{11}$ & $t_{12}$ & $t_{13}$ & $t_{14}$ & $t_{15}$ & $t_{16}$ & $t_{17}$ & $t_{18}$ & $t_{19}$ & $t_{20}$ & $t_{21}$ \\
         \hline \hline
         $v_1$ & $1$ & $1$ & $1$ & $1$ & $1$ & $1$ & $1$ & $1$ & $1$ & $1$ & $1$ & $\cellcolor{red!25}-1$ & $\cellcolor{red!25}-1$ & $\cellcolor{red!25}-1$ & $\cellcolor{red!25}-1$ & $\cellcolor{red!25}-1$ & $\cellcolor{red!25}-1$ & $\cellcolor{red!25}-1$ & $\cellcolor{red!25}-1$ & $\cellcolor{red!25}-1$ & $\cellcolor{red!25}-1$ \\
         \hline
         $v_2$ & $1$ & $1$ & $1$ & $1$ & $1$ & $1$ & $1$ & $1$ & $1$ & $1$ & $1$ & $\cellcolor{red!25}-1$ & $\cellcolor{red!25}-1$ & $\cellcolor{red!25}-1$ & $\cellcolor{red!25}-1$ & $\cellcolor{red!25}-1$ & $\cellcolor{red!25}-1$ & $\cellcolor{red!25}-1$ & $\cellcolor{red!25}-1$ & $\cellcolor{red!25}-1$ & $\cellcolor{red!25}-1$ \\
         \hline
         $v_3$ & $1$ & $\cellcolor{red!25}-1$ & $\cellcolor{red!25}-1$ & $\cellcolor{red!25}-1$ & $\cellcolor{red!25}-1$ & $\cellcolor{red!25}-1$ & $\cellcolor{red!25}-1$ & $1$ & $1$ & $1$ & $1$ & $\cellcolor{red!25}-1$ & $\cellcolor{red!25}-1$ & $\cellcolor{red!25}-1$ & $\cellcolor{red!25}-1$ & $\cellcolor{red!25}-1$ & $\cellcolor{red!25}-1$ & $\cellcolor{red!25}-1$ & $\cellcolor{red!25}-1$ & $\cellcolor{red!25}-1$ & $\cellcolor{red!25}-1$ \\
         \hline
         $v_4$ & $1$ & $1$ & $\cellcolor{red!25}-1$ & $\cellcolor{red!25}-1$ & $\cellcolor{red!25}-1$ & $\cellcolor{red!25}-1$ & $\cellcolor{red!25}-1$ & $\cellcolor{red!25}-1$ & $1$ & $1$ & $1$ & $1$ & $1$ & $1$ & $1$ & $1$ & $1$ & $\cellcolor{red!25}-1$ & $\cellcolor{red!25}-1$ & $\cellcolor{red!25}-1$ & $\cellcolor{red!25}-1$ \\
         \hline
         $v_5$ & $1$ & $1$ & $1$ & $\cellcolor{red!25}-1$ & $\cellcolor{red!25}-1$ & $\cellcolor{red!25}-1$ & $\cellcolor{red!25}-1$ & $\cellcolor{red!25}-1$ & $\cellcolor{red!25}-1$ & $1$ & $1$ & $\cellcolor{red!25}-1$ & $1$ & $1$ & $1$ & $1$ & $1$ & $1$ & $\cellcolor{red!25}-1$ & $\cellcolor{red!25}-1$ & $\cellcolor{red!25}-1$ \\
         \hline
         $v_6$ & $1$ & $1$ & $1$ & $1$ & $\cellcolor{red!25}-1$ & $\cellcolor{red!25}-1$ & $\cellcolor{red!25}-1$ & $\cellcolor{red!25}-1$ & $\cellcolor{red!25}-1$ & $\cellcolor{red!25}-1$ & $1$ & $\cellcolor{red!25}-1$ & $\cellcolor{red!25}-1$ & $1$ & $1$ & $1$ & $1$ & $1$ & $1$ & $\cellcolor{red!25}-1$ & $\cellcolor{red!25}-1$ \\
         \hline
         $v_7$ & $1$ & $1$ & $1$ & $1$ & $1$ & $\cellcolor{red!25}-1$ & $\cellcolor{red!25}-1$ & $\cellcolor{red!25}-1$ & $\cellcolor{red!25}-1$ & $\cellcolor{red!25}-1$ & $\cellcolor{red!25}-1$ & $\cellcolor{red!25}-1$ & $\cellcolor{red!25}-1$ & $\cellcolor{red!25}-1$ & $1$ & $1$ & $1$ & $1$ & $1$ & $1$ & $\cellcolor{red!25}-1$ \\
         \hline
         $v_8$ & $\cellcolor{red!25}-1$ & $1$ & $1$ & $1$ & $1$ & $1$ & $\cellcolor{red!25}-1$ & $\cellcolor{red!25}-1$ & $\cellcolor{red!25}-1$ & $\cellcolor{red!25}-1$ & $\cellcolor{red!25}-1$ & $\cellcolor{red!25}-1$ & $\cellcolor{red!25}-1$ & $\cellcolor{red!25}-1$ & $\cellcolor{red!25}-1$ & $1$ & $1$ & $1$ & $1$ & $1$ & $1$ \\
         \hline
         $v_9$ & $\cellcolor{red!25}-1$ & $\cellcolor{red!25}-1$ & $1$ & $1$ & $1$ & $1$ & $1$ & $\cellcolor{red!25}-1$ & $\cellcolor{red!25}-1$ & $\cellcolor{red!25}-1$ & $\cellcolor{red!25}-1$ & $1$ & $\cellcolor{red!25}-1$ & $\cellcolor{red!25}-1$ & $\cellcolor{red!25}-1$ & $\cellcolor{red!25}-1$ & $1$ & $1$ & $1$ & $1$ & $1$ \\
         \hline
         $v_{10}$ & $\cellcolor{red!25}-1$ & $\cellcolor{red!25}-1$ & $\cellcolor{red!25}-1$ & $1$ & $1$ & $1$ & $1$ & $1$ & $\cellcolor{red!25}-1$ & $\cellcolor{red!25}-1$ & $\cellcolor{red!25}-1$ & $1$ & $1$ & $\cellcolor{red!25}-1$ & $\cellcolor{red!25}-1$ & $\cellcolor{red!25}-1$ & $\cellcolor{red!25}-1$ & $1$ & $1$ & $1$ & $1$ \\
         \hline
         $v_{11}$ & $\cellcolor{red!25}-1$ & $\cellcolor{red!25}-1$ & $\cellcolor{red!25}-1$ & $\cellcolor{red!25}-1$ & $1$ & $1$ & $1$ & $1$ & $1$ & $\cellcolor{red!25}-1$ & $\cellcolor{red!25}-1$ & $1$ & $1$ & $1$ & $\cellcolor{red!25}-1$ & $\cellcolor{red!25}-1$ & $\cellcolor{red!25}-1$ & $\cellcolor{red!25}-1$ & $1$ & $1$ & $1$ \\
         \hline
         $v_{12}$ & $\cellcolor{red!25}-1$ & $\cellcolor{red!25}-1$ & $\cellcolor{red!25}-1$ & $\cellcolor{red!25}-1$ & $\cellcolor{red!25}-1$ & $1$ & $1$ & $1$ & $1$ & $1$ & $\cellcolor{red!25}-1$ & $1$ & $1$ & $1$ & $1$ & $\cellcolor{red!25}-1$ & $\cellcolor{red!25}-1$ & $\cellcolor{red!25}-1$ & $\cellcolor{red!25}-1$ & $1$ & $1$ \\
         \hline
         $v_{13}$ & $\cellcolor{red!25}-1$ & $\cellcolor{red!25}-1$ & $\cellcolor{red!25}-1$ & $\cellcolor{red!25}-1$ & $\cellcolor{red!25}-1$ & $\cellcolor{red!25}-1$ & $1$ & $1$ & $1$ & $1$ & $1$ & $1$ & $1$ & $1$ & $1$ & $1$ & $\cellcolor{red!25}-1$ & $\cellcolor{red!25}-1$ & $\cellcolor{red!25}-1$ & $\cellcolor{red!25}-1$ & $1$ \\
         \hline
    \end{tabular}
    \caption{A large bundle of 21 issues for 13 voters. The first 11 issues form the majority-supported subbundle and the remaining have only minority support. However, the first subbundle has only two supporters while the second has support from all but three voters.}
    \label{tab:profile3}
\end{table}

Table \ref{tab:profile3} shows how dividing the bundle into majority- and minority-supported subbundles can achieve the opposite of the intended result. This set of voters has a total of 273 voter-issue pairs, and almost exactly half (137) are positive, while the other half (136) are negative. This suggests there should be a number of popular and unpopular issues in this bundle. Indeed, the first 11 of the 21 issues have majority support and the last 10 do not. However, when we divide the bundle into the majority and minority supported issues, we make things worse, not better! The majority support subbundle fails with only two supporters, and the minority support subbundle passes with only three dissenters. The resulting subbundle vote has an issue score of -1, the worst possible outcome! In pizza terms, the group agreed to all the bad toppings and rejected all the good ones.

Applying a similar method to the bounds in Equations \eqref{eq:bound1} and \eqref{eq:bound2} (see Supplementary Information for derivations), we can determine how many supporters the majority subbundle \emph{must} have and the maximum number of supporters the minority subbundle \emph{can} have.

\begin{equation}
    \text{Number of voters supporting majority subbundle} \geq \ceil*{
\frac{m_1+nc}{m_1+c}
}.
\end{equation}

\begin{equation}
    \text{Number of voters supporting minority subbundle} \leq \floor*{
\frac{m_2(n-1)}{m_2+c}
}.
\end{equation}

Here, $c=1$ for the subbundle with an odd number of issues and $c=2$ for the subbundle with an even number. Like before, these bounds are not very restrictive. In the voting profile in Table \ref{tab:profile3}, we have $n = 13$, $m_1 = 11$, and $m_2 = 10$ and these bounds are equalities, so we know that these subbundles are maximally manipulated. 

There are still strong strategic considerations when deciding to subdivide a bundle. In extreme cases, a bundle that is going to pass is divided into two subbundles, both of which fail (or vice versa). A pizza metaphor would be the following example: A group consists of three herbivores, three carnivores, and one omnivore. The herbivores are turned off by the meats on the supreme pizza and the carnivores are turned off by the veggies, so the group chooses to get plain cheese. The pizza seller, disappointed by the lack of topping sales, first asks if the group would like to add just the meat toppings. The carnivore and omnivore coalition of the group agrees to this and the vegetarians are outvoted. Then the group is asked if they would like to add just the veggie toppings and they agree again, this time the herbivores allying with the omnivore. By making it two choices instead of one, the groups decision on every topping has been reversed. An example of this is shown in Table \ref{tab:profile4}. Bundles like this example are easy to come by; the probabilities of this occurring to a random bundle division can be seen for various $n$ and $m$ can be seen in Table \ref{tab:split_percentages}, ranging from 2.5\% to 8.5\%. 

\begin{table}
    \centering
    \small
    \begin{tabular}{m{0.25cm}||m{0.2cm}|m{0.2cm}|m{0.2cm}|m{0.2cm}|m{0.2cm}|m{0.2cm}|m{0.2cm}||m{0.2cm}|m{0.2cm}|m{0.2cm}|m{0.2cm}|m{0.2cm}|m{0.2cm}|m{0.2cm}|m{0.2cm}|}
         & $t_1$ & $t_2$ & $t_3$ & $t_4$ & $t_5$ & $t_6$ & $t_7$ & $t_8$ & $t_9$ & $t_{10}$ & $t_{11}$ & $t_{12}$ & $t_{13}$ & $t_{14}$ & $t_{15}$
         \\
         \hline \hline
         $v_1$ & $\cellcolor{red!25}-1$ & $\cellcolor{red!25}-1$ & $\cellcolor{red!25}-1$ & $\cellcolor{red!25}-1$ & $\cellcolor{red!25}-1$ & $\cellcolor{red!25}-1$ & $\cellcolor{red!25}-1$ & $\cellcolor{red!25}-1$ & $1$ & $1$ & $1$ & $1$ & $1$ & $1$ & $1$ \\
         \hline
         $v_2$ & $1$ & $1$ & $1$ & $1$ & $1$ & $1$ & $1$ & $\cellcolor{red!25}-1$ & $\cellcolor{red!25}-1$ & $\cellcolor{red!25}-1$ & $\cellcolor{red!25}-1$ & $\cellcolor{red!25}-1$ & $\cellcolor{red!25}-1$ & $\cellcolor{red!25}-1$ & $\cellcolor{red!25}-1$ \\
         \hline
         $v_3$ & $\cellcolor{red!25}-1$ & $\cellcolor{red!25}-1$ & $\cellcolor{red!25}-1$ & $\cellcolor{red!25}-1$ & $\cellcolor{red!25}-1$ & $\cellcolor{red!25}-1$ & $\cellcolor{red!25}-1$ & $1$ & $\cellcolor{red!25}-1$ & $1$ & $1$ & $1$ & $1$ & $1$ & $1$ \\
         \hline
         $v_4$ & $1$ & $1$ & $1$ & $1$ & $1$ & $1$ & $1$ & $\cellcolor{red!25}-1$ & $\cellcolor{red!25}-1$ & $\cellcolor{red!25}-1$ & $\cellcolor{red!25}-1$ & $\cellcolor{red!25}-1$ & $\cellcolor{red!25}-1$ & $\cellcolor{red!25}-1$ & $\cellcolor{red!25}-1$ \\
         \hline
         $v_5$ & $\cellcolor{red!25}-1$ & $\cellcolor{red!25}-1$ & $\cellcolor{red!25}-1$ & $\cellcolor{red!25}-1$ & $\cellcolor{red!25}-1$ & $\cellcolor{red!25}-1$ & $\cellcolor{red!25}-1$ & $1$ & $1$ & $\cellcolor{red!25}-1$ & $1$ & $1$ & $1$ & $1$ & $1$ \\
         \hline
         $v_6$ & $1$ & $1$ & $1$ & $1$ & $1$ & $1$ & $1$ & $\cellcolor{red!25}-1$ & $\cellcolor{red!25}-1$ & $\cellcolor{red!25}-1$ & $\cellcolor{red!25}-1$ & $\cellcolor{red!25}-1$ & $\cellcolor{red!25}-1$ & $\cellcolor{red!25}-1$ & $\cellcolor{red!25}-1$ \\
         \hline
         $v_7$ & $1$ & $1$ & $1$ & $1$ & $1$ & $1$ & $1$ & $1$ & $1$ & $1$ & $1$ & $1$ & $1$ & $1$ & $1$ \\
         \hline
    \end{tabular}
    \caption{A bundle that is rejected by the group 1-6, but when divided into two subbundles, both pass 4-3.}
    \label{tab:profile4}
\end{table}

\begin{table}[]
    \centering
    \begin{tabular}{r||c|c|c|}
         & $m=11$ & $m=101$ & $m=1001$ \\
         \hline \hline
        $n=11$ & 3.3 (1.5) & 5.7 (2.3) & 5.9 (2.5) \\
        \hline
        $n=101$ & 5.6 (2.1) & 7.0 (2.1) & 7.7 (2.6) \\
        \hline
        $n=1001$ & 5.8 (2.3) & 6.9 (2.8) & 8.6 (2.6) \\
        \hline
    \end{tabular}
    \caption{The percent of IID utility profiles where the bundled vote is completely reversed when randomly divided into two subbundles. A bootstrapping method with 100 iterations of 100 samples was used for each pair of $n$ and $m$, with standard deviation shown in parentheses.}
    \label{tab:split_percentages}
\end{table}

\subsection{Subbundling Schemes of Many Subbundles}

The reason bundling is less effective than issue-by-issue voting is that voters are typically forced to vote for issues they do not like to gain a vote for the marginally larger number of issues they do support (or vice versa). The larger a bundle is, the more issues a voter will be forced to vote for (or against), in spite of their preferences on that particular issue. Therefore, it stands to reason that as we divide a bundle into smaller bundles, the expected value from the sum of the subbundle votes should be greater than the value of the bundle. After all, the issue-by-issue vote, which has maximal value in both utility and issue scores, can be considered a vote on $m$ single-issue bundles.

\begin{figure}
    \centering
    \includegraphics[width=\textwidth]{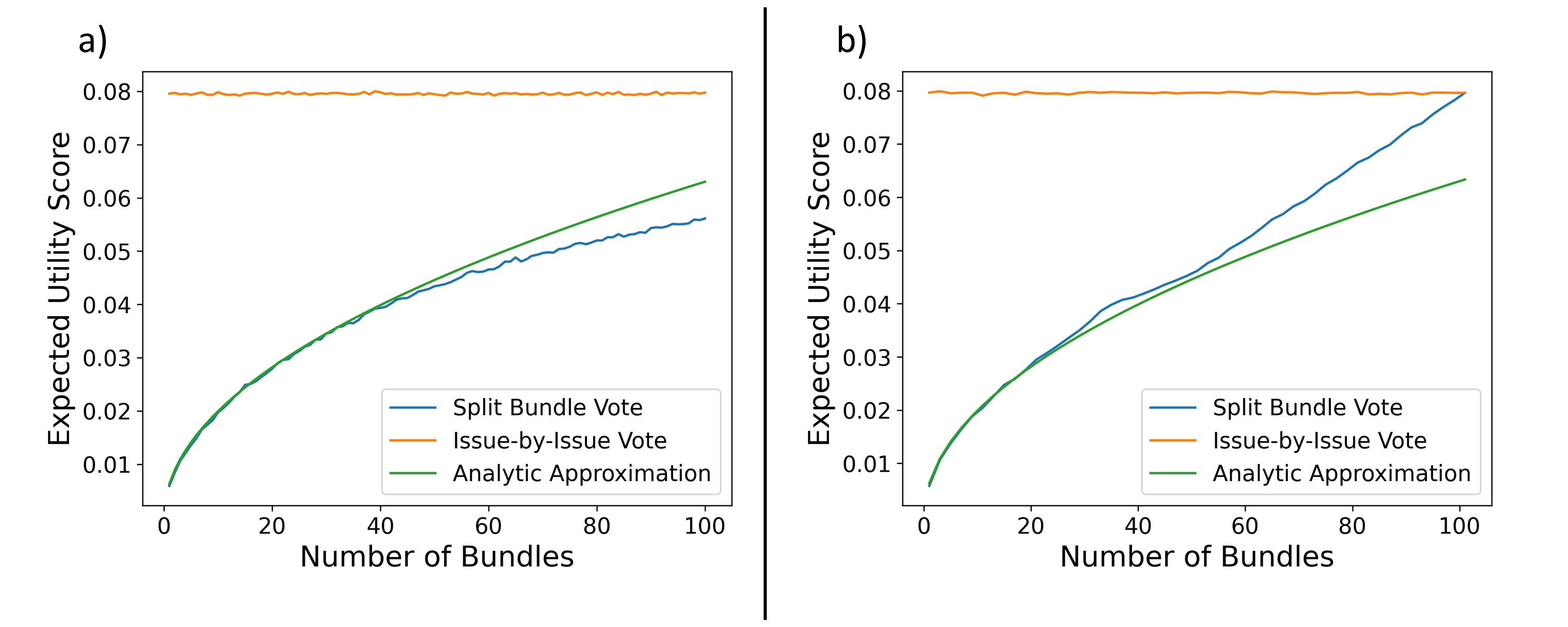}
    \caption{The utility scores for subbundle votes as a function of how many bundles were used. In (a), issues are randomly assigned to one of $k$ subbundles. The utility score of the split bundle vote does not reach the value of issue-by-issue voting, even when $k=m$, because many subbundles have no issues while other have many. In (b), subbundles are created ensuring that each subbundle is roughly the same size (while also making sure each subbundle has an odd number of issues for tie-breakers). For example, if we have 21 issues into 5 subbundles, instead of creating subbundles of size 5, 4, 4, 4, and 4, where tie-breaking effects would be substantial, we create them with sizes 5, 5, 5, 3, and 3. When subbundles are created this way, we get much more value when the number of subbundles is large compared to the purely random method, since there are no wasted subbundles with zero issues.}
    \label{fig:subbundles}
\end{figure}

Equation \eqref{eq:analytic_score} shows that if we divide a bundle into $k$ subbundles (randomly, not according to an agenda-setter), the expected utility score increases by $\frac{k}{\sqrt{k}}=\sqrt{k}$, so there is a law of diminishing returns on the number of subbundles. This approximation breaks down as the subbundles quickly get small until $k = m$, when this becomes simple issue-by-issue voting. We see the results of simulations and the corresponding analytic approximations in Figure \ref{fig:subbundles}. The main takeaway is that \emph{random} division of a bundle into subbundles results in immediate added value for the group. We have shown (and will continue to show) that an NVAAS can find anti-democratic bundles and subbundles, but these are rare, and the average bundle division has positive value. Therefore, when the bundle division is done randomly, and not by a potentially malevolent actor, there is value to be gained that only needs to be weighed against the cost of additional votes.

We have shown that when utilities are random, bundles tend to be well-behaved with positive value on average, but can be counterproductive when manipulated by an NVAAS. Our previous examples of manipulation in Tables \ref{tab:profile2}, \ref{tab:profile3}, and \ref{tab:profile4} have been carefully constructed by placing $1$s and $-1$s. However, this last result suggests that, with surprising frequency, even uniformly generated bundles in the average case can be manipulated to achieve anti-majoritarian goals by dividing them up into subbundles. This is a significant shift, showing manipulation is not just \emph{possible} but (relatively) \emph{easy}.

To study the potential for ordinary bundles to be abused by bad actors, we create utility profiles using the IID utility model and then examine each possible subbundle. Using the efficiency gap score from gerrymandering \cite{duchin_2018}, we evaluate the reasonableness of that subbundle. We also compute the issue score, our bespoke measure designed specifically for multi-issue group decision-making.

\begin{figure}
    \centering
    \includegraphics[width=\textwidth]{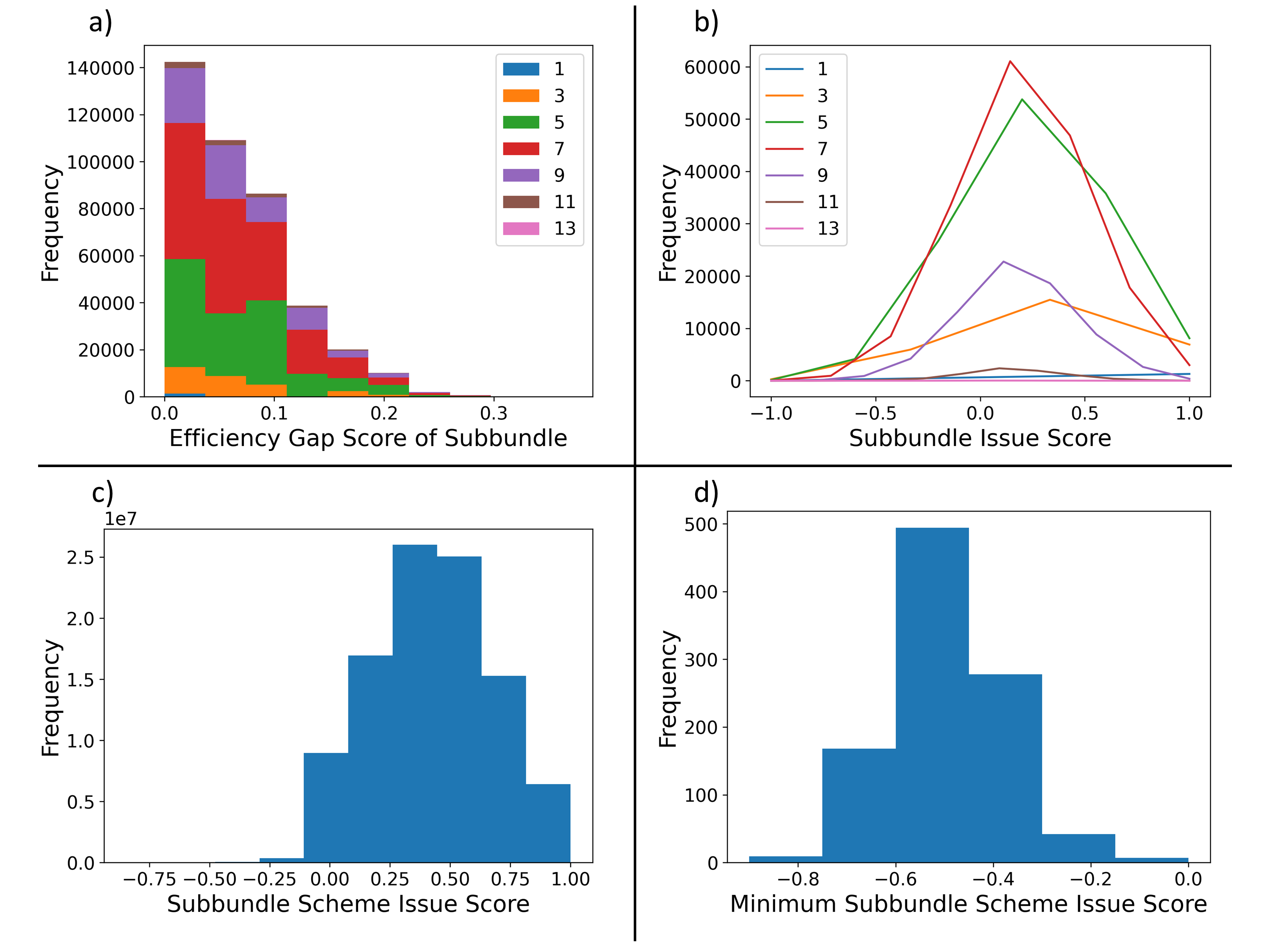}
    \caption{Prevalence of potential manipulation in an average utility profile. In (a), we see that most subbundles are well-behaved according to the efficiency gap metric. However, this is not the whole story, as we can see in (b) that many subbundles have very low issue scores, suggesting anti-democratic outcomes. In (a) and (b) results are sorted by the size of the subbundle.
    Then we add the additional restraint that each issue must appear in one of the subbundles in a scheme. (c) shows us the distribution of issue scores for subbundle schemes. The vast majority have positive scores, and it is difficult to see how low the issue scores can go. (d) solves this problem by showing the distribution of the lowest possible issue scores, and we see that most utility profiles can be significantly manipulated, with around $\frac{3}{4}$ of issues having anti-democratic outcomes, and potentially almost all issues.}
    \label{fig:subbundle_gerry}
\end{figure}


Again, the average case is promising. For reference, the subbundles in Table \ref{tab:profile3} have efficiency gap scores of 0.45 and 0.4 and issue scores of -1. A typical subbundle has a a small efficiency gap (Figure \ref{fig:subbundle_gerry}a), and a positive issue score (Figure \ref{fig:subbundle_gerry}b). Just like large bundles, random subbundles still generate positive value. However, inside every large bundle, there are specific subbundles that NVAASes could call up for a vote in order to pass a significant minority agenda. Many subbundles have large negative issue scores. 

This susceptibility to manipulation persists even if we require that all issues must be voted on and included in a subbundle. In Figure \ref{fig:subbundle_gerry}c-d, we analyze the ability of an individual with minority views on all issues to create a \emph{subbundle scheme} by dividing a bundle into any number of subbundles to maximize the number of issues passed or failed against the majority. Beginning with a random utility profile, we consider every possible subbundle scheme that uses only odd sized subbundles (to avoid tiebreaker effects). The most extreme case seen in Figure \ref{fig:subbundle_gerry}d is a subbundling scheme in which 12 of the 13 issues are passed/failed against the majority, for a total issue score of $\frac{-11}{13}$. Almost all utility profiles have a subbundle scheme in which more than $\frac{2}{3}$ of all issues are passed or failed against the majority. This is on par with the extreme case bounds that we derived earlier, suggesting that the majority of bundles have the potential for strategic manipulation on the same level as the most extreme bundles. 

Returning one last time to the pizza example, suppose the owner has 13 toppings, none of which are liked by the majority of the customers. By carefully bundling ingredients, the group will agree to almost all of the ingredients. And this can be done with almost any group (although a new subbundling scheme will be needed for each group), not just very particular groups like the one in Table \ref{tab:profile4}.

\subsection{A Realistic Scenario}

To demonstrate the effectiveness of this method, we create a hypothetical committee of lawmakers using real American voters  and test how well an NVAAS could manipulate the group into an undemocratic result. Our data for individual preferences comes from the ANES's 2020 time series data, specifically the post-election survey \cite{anes_2020}. There were several phases of processing the data to get to our final result, which we describe here.

This survey included hundreds of questions, including 15 questions that revealed an respondent's utility for an legislative issue. For computational reasons (an exhaustive search of all possible subbundle schemes gets computationally expensive quickly as $m$ increases), we reduced this set to 11 issues by removing the 4 issues that had the most consensus. The details of the issues we used, including their ANES variable codes, can be found in the SI. We collected the responses and created a large utility profile of all the respondents. However, each question allowed a neutral response (neither support or opposition). Since our model does not allow utility 0, we discarded all respondents who were indifferent to one of the 11 issues. The initial sample of voters was fairly evenly distributed across party identification (2318 Democrats, 2250 Republicans, and 3712 other). After filtering out the indifferent respondents, however, our population was more imbalanced, with 436 Democrats, 130 Republicans, and 278 other. To restore balance to our group, have a more general statement about group utilities, and also reduce the population for computational reasons, we sampled from this larger group to create small committees of 10 Democrats, 10 Republicans, and 11 others. For each of these 31 voter groups, we examine all possible subbundle schemes to find the one most beneficial to an NVAAS. 

\begin{figure}
    \centering
    \includegraphics[width=\textwidth]{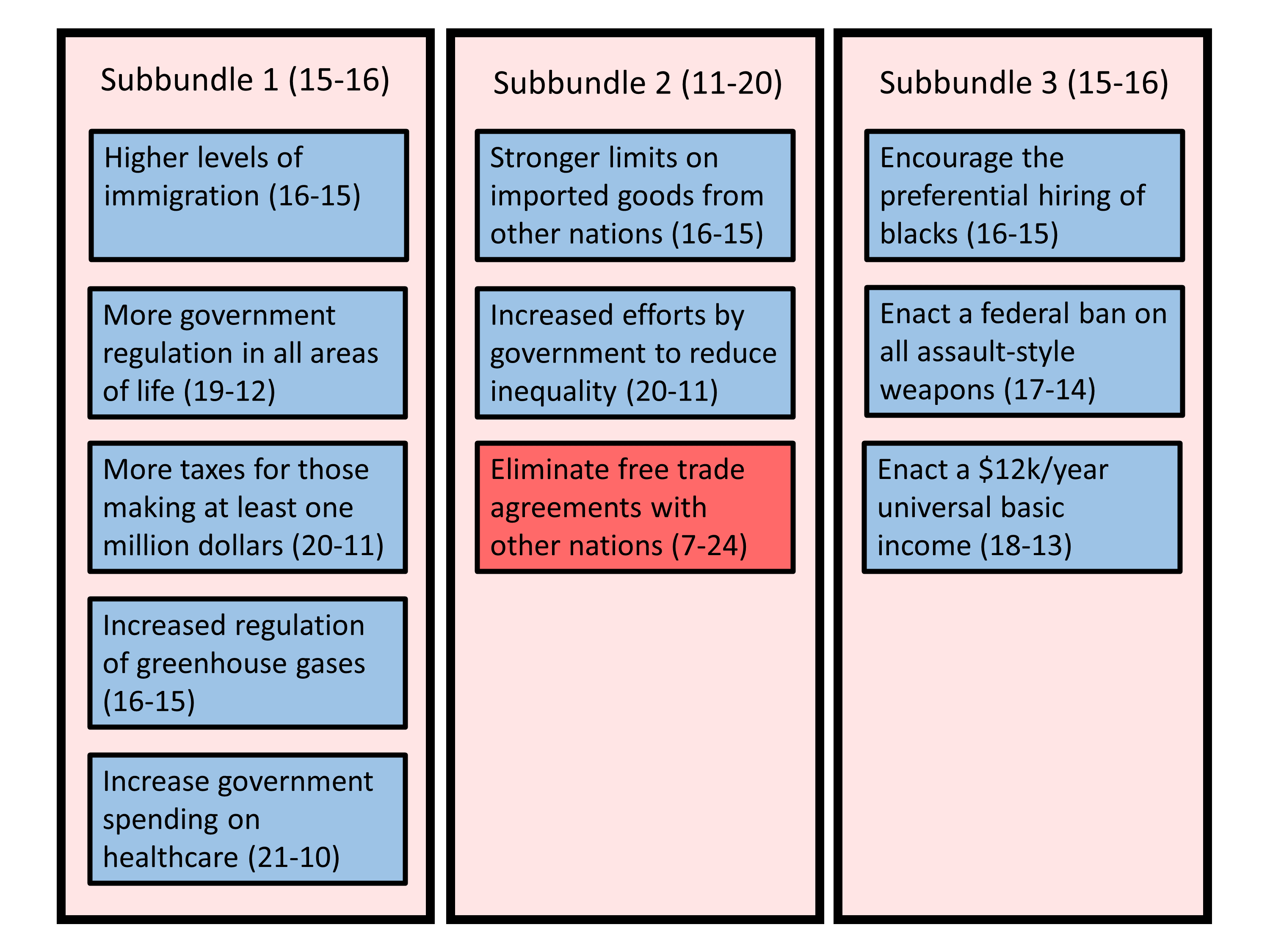}
    \caption{A subbundling scheme for 11 issues and 31 voters. Color represents if the subbundle/issue is preferred (blue) or opposed (red) by a majority, and the vote result on each subbundle/issue is shown in parentheses.  This is the most heavily manipulated subbundle scheme found from the ANES data, with the majority only happy about the outcome on one issue (free trade agreements). On the rest, the NVAAS was able to achieve the minority preference by rejecting popular issues.}
    \label{fig:max_gerry}
\end{figure}

All 1,000 simulated committees had a subbundle scheme with a negative issue score. Most had scores of $\frac{-5}{11}$ or $\frac{-3}{11}$, meaning the NVAAS was successful on 8 or 7 issues, respectively. One group, shown in Figure \ref{fig:max_gerry}, had an issue score of $\frac{-9}{11}$ in which only one of the 11 issues had a majority-preferred outcome! The full distribution of minimum issue scores, as well as a more representative subbundling scheme, can be found in the Supplementary Information.

This example showcases the power of antagonistic subbundling to subvert the will of the majority, even when the group's utility profile has an underlying structure as individual preferences are driven by political orientation.

\section{Discussion and Conclusion}

The benefits and costs of bundling a large number of decisions into a single vote to simplify the group decision process can be hard to assess. The group is saved from having to query its members $m$ times and instead gets to hold a single vote. The group members only have to make one determination of value instead of $m$ separate choices. However, this economical decision-making comes at a cost. By reducing to a single yes/no vote, members lose the ability to draw distinctions between issues, which typically results in an outcome with a lower utility for the group than issue-by-issue voting. This method also leaves the group vulnerable to manipulation by an NVAAS, as these issues can usually be bundled in very specific ways to achieve anti-democratic ends in which most issues are accepted or rejected against the will of the majority. 

If the preferences of each individual on each issue (a complete utility profile) is collected, that information should be used to automatically carry out issue-by-issue voting. Random bundling is a benign process that can genuinely save time and energy for the entire group, but if an agent is given that information and then allowed to set the agenda by creating subbundles, the result will largely be aligned with the utility of that agent, not the group as a whole. This information is at least partially known in many situations: congressional committee chairs likely know a great deal about the preferences of the members of their committee and for-profit companies spend a great deal of money analyzing consumer preferences and trends. Thus, most groups are potentially vulnerable to this manipulation, regardless of any underlying structure in the utility profile.

\subsection{Assumptions and Consequences}

As mentioned earlier, this model makes two assumptions about human preferences: separability and equal utility. These are strong assumptions, so we take some extra time to define and justify these assumptions, as well as discuss the implications of these conditions.

The first assumption is separability, which means voter preference on one issue is not dependent on the success or failure of any other issue \cite{lang_xia_2016, lacy_2001}. This condition is essential for studying issue-by-issue voting; if a voter only prefers ham on a pizza if there is already pineapple on the pizza, there is no clean, principled way to determine how she should vote on ham when the outcome of the pineapple vote is unknown. Voter preferences are separable more often than not, even on related issues like abortion access and Aid to Families with Dependent Children programs \cite{lacy_2001}. Removing this assumption would overburden our model, require further assumptions about voter behavior, and obscure 
our results even further. We would need to take into account voters who play it safe (voting no on ham just in case pineapple is not approved) or who risk it all (vote yes for ham in the hopes that pineapple will also be approved) to every type of behavior in the middle. This is a valuable area for future research.

The second assumption, that all issues are equally important to each voter, is necessary to remove the temptation for voters to trade votes. Variation in issue utility leads to self-interested voters giving away votes on low-utility issues in exchange for votes on high-utility issues which opens up all sorts of interesting dynamics \cite{xefteris_ziros_2017, riker_brams_1973, casella_palfrey_2019}. The precise way in which bundling interacts with vote trading is an open question and is also a natural direction for future work.

These assumptions also simplify other aspects of the multi-issue decision-making process. 

\begin{itemize}
    \item There are no single-issue voters \cite{congleton_1991} because all issues have equal weight. 
    \item We avoid the issue of exernalities by incorporating any ``unintended consequences'' in the utilities of the issues. For example, if one member of the group really likes sausage but knows that their vegan friend will not be able to eat it, they may have a $-1$ for that utility despite personally liking it.
    \item Brinksmanship on large issues like the U.S. debt ceiling  \cite{herra_mace_nunez_2022} appears when groups cannot decide which issues should be bundled together. Since we implicitly assume that a bundle has already been created, we do not have any brinksmanship concerns. See also the following point.
    \item Because issues are separable, voting behavior is independent of the sequence in which issue-by-issue votes are cast. Therefore, we can assume they are simultaneous. 
    \item With rational voters making binary decisions, simple majority rule satisfies all the conditions for sincere voting \cite{arrow_1963}, and there is no reason for a voter to vote for a single issue/bundle that they oppose or vice versa.
    \item We assume that the bundle in question is competing against the status quo. If in fact it is two separate bundles that voters are deciding between, because issues are separable, we can look only at the issues in which the two bundles differ. One bundle can be labelled as the status quo and let voters decide to pass or fail the other bundle, so our analysis here holds even for competition between bundles with no status quo.
    \item In this work, we completely ignore questions of information and information asymmetry between voters. This is because voting is sincere, and therefore a voter's actions do not depend on what she perceives other voters' intentions to be. However, an NVAAS must have total information in order to successfully manipulate the group.
\end{itemize}

Many of these things result in new and interesting dynamics, and this section provides many future directions of exploration by loosening specific assumptions to see how rational behavior changes.

\subsection{Legislative Implications}

This paper describes mathematically how majority rule decision-making processes for bundled decisions may result in drastically different outcomes than majority rule decision-making processes without bundling. In a legislative context, where bundled legislation is the norm and dividing legislation is difficult, our findings imply that when legislation passes by majority vote, thereby achieving a theoretically positive aggregate utility, it is mathematically possible that nearly all of the individual issues within those legislative bundles are unsupported by a majority of legislators. Majority rule by bundling and majority rule on individual issues could lead to radically different legislative outcomes. This result implies that a high proportion of existing law, were it divided into component measures, would not be supported by any majority. The study further shows that partially dividing legislation cannot solve this problem.

These findings raise fundamental questions about the design of legislative processes. Should the goal of a legislature be to maximize the aggregate utility of its legislators, or to pass legislation that garners majority support? What is the role of bundling in the theoretical and practical pursuit of those goals? By optimizing for aggregate utility, lawmakers get more of what they want. However, this comes at the expense of passing a large body of law that a majority of members do not want. Optimizing instead for majority rule by issue, notwithstanding practicality concerns, would likely reduce the amount of law created, but would ensure that resulting law was backed by true legislative majorities. This method would preserve the majority rule principle that many, perhaps falsely, believe is reflected in current practice but might alienate minorities that receive fewer benefits from the legislative process. Additionally, legislatures deal with highly nonseparable issues. Conflict can also arise if, for example, a legislature votes to increase the number of police or teachers but refuses to increase taxes to pay for them in a separate vote. 

Beyond the disconnect in majority rule between bundled and unbundled legislation, the mathematics presented here show that NVAASes with asymmetric information advantages could design legislation that maximizes or minimizes either the aggregate utility or the number of underlying issues supported by a majority without any apparent difference in the final vote tally. Lower cost legislation, that is, legislation composed of issues with fewer supporters, could be favored by bundlers seeking to maximize their own utility. A question to be explored in future work addresses what information sharing or legislative processes might mitigate this hazard. One option is voting reforms. There is evidence to suggest that other voting methods like approval voting \cite{Brams_1995} or yes-no voting \cite{Brams_Fishburn_1993} are less susceptible to bundling-related manipulation than simple plurality voting.

\bibliography{refs}

\begin{thebibliography}{10}

\bibitem{hillinger_1971}
Claude Hillinger.
\newblock Voting on issues and on platforms.
\newblock {\em Behavioral Science}, 16(6):564–566, 1971.

\bibitem{kadane_1972}
Joseph~B. Kadane.
\newblock On division of the question.
\newblock {\em Public Choice}, 13(1):47–54, 1972.

\bibitem{RAE_DAUDT_1976}
Douglas~W. Rae and Hans Daudt.
\newblock The ostrogorski paradox: A peculiarity of compound majority decision
  *.
\newblock {\em European Journal of Political Research}, 4(4):391–398, 1976.

\bibitem{Wagner_1983}
Carl Wagner.
\newblock Anscombe’s paradox and the rule of three-fourths.
\newblock {\em Theory and Decision}, 15(3):303–308, 1983.

\bibitem{Brams_Kilgour_Zwicker_1998}
Steven~J. Brams, D.~Marc Kilgour, and William~S. Zwicker.
\newblock The paradox of multiple elections.
\newblock {\em Social Choice and Welfare}, 15(2):211–236, 1998.

\bibitem{Gehrlein_Merlin_2020}
William~V. Gehrlein and Vincent Merlin.
\newblock On the probability of the ostrogorski paradox.
\newblock {\em Studies in Choice and Welfare}, page 119–135, 2020.

\bibitem{Riker_1986}
William~H. Riker.
\newblock {\em The art of political manipulation}.
\newblock Yale Univ. Pr., 1986.

\bibitem{krutz_2001}
Glen~S. Krutz.
\newblock {\em Hitching a ride: Omnibus legislating in the US congress}.
\newblock Ohio State University Press, 2001.

\bibitem{krutz_2021}
Glen~S. Krutz.
\newblock Omnibus legislating in the u.s. congress.
\newblock {\em Legisprudence Library}, page 35–51, 2021.

\bibitem{rundquist_strom_1987}
Barry~S. Rundquist and Gerald~S. Strom.
\newblock Bill construction in legislative committees: A study of the u. s.
  house.
\newblock {\em Legislative Studies Quarterly}, 12(1):97, 1987.

\bibitem{milkman_mazza_shu_tsay_bazerman_2012}
Katherine~L. Milkman, Mary~Carol Mazza, Lisa~L. Shu, Chia-Jung Tsay, and Max~H.
  Bazerman.
\newblock Policy bundling to overcome loss aversion: A method for improving
  legislative outcomes.
\newblock {\em Organizational Behavior and Human Decision Processes},
  117(1):158–167, 2012.

\bibitem{volokh_2011}
Hanah~Metchis Volokh.
\newblock Read-the-bill rule for congress, a.
\newblock {\em Missouri Law Review}, 76(1):135–184, 2011.

\bibitem{congleton_1991}
Roger~D. Congleton.
\newblock Information, special interests, and single-issue voting.
\newblock {\em Public Choice}, 69(1):39–49, 1991.

\bibitem{alonso_camara_2016}
Ricardo Alonso and Odilon C\^{a}mara.
\newblock Persuading voters.
\newblock {\em American Economic Review}, 106(11):3590–3605, 2016.

\bibitem{camara_eguia_2017}
Odilon C\^{a}mara and Jon~X. Eguia.
\newblock Slicing and bundling.
\newblock {\em The Journal of Politics}, 79(4):1460–1464, 2017.

\bibitem{gabbe_pierce_2016}
C.~J. Gabbe and Gregory Pierce.
\newblock Hidden costs and deadweight losses: Bundled parking and residential
  rents in the metropolitan united states.
\newblock {\em Housing Policy Debate}, 27(2):217–229, 2016.

\bibitem{doyle_2018}
Kevin Doyle.
\newblock Sticker shock: Navigating car trim levels, Aug 2018.

\bibitem{priessner_hampl_2020}
Alfons Priessner and Nina Hampl.
\newblock Can product bundling increase the joint adoption of electric
  vehicles, solar panels and battery storage? explorative evidence from a
  choice-based conjoint study in austria.
\newblock {\em Ecological Economics}, 167:106381, 2020.

\bibitem{heeler_nguyen_buff_2007}
Roger~M. Heeler, Adam Nguyen, and Cheryl Buff.
\newblock Bundles = discount? revisiting complex theories of bundle effects.
\newblock {\em Journal of Product \& Brand Management}, 16(7):492–500, 2007.

\bibitem{xefteris_ziros_2017}
Dimitrios Xefteris and Nicholas Ziros.
\newblock Strategic vote trading in power sharing systems.
\newblock {\em American Economic Journal: Microeconomics}, 9(2):76–94, 2017.

\bibitem{herrera_morelli_palfrey_2014}
Helios Herrera, Massimo Morelli, and Thomas Palfrey.
\newblock Turnout and power sharing.
\newblock {\em The Economic Journal}, 124(574), 2014.

\bibitem{jones_sirianni_fu_2022}
Matthew~I. Jones, Antonio~D. Sirianni, and Feng Fu.
\newblock Polarization, abstention, and the median voter theorem.
\newblock {\em Humanities and Social Sciences Communications}, 9(1), 2022.

\bibitem{poole_rosenthal_1997}
Keith~T. Poole and Howard Rosenthal.
\newblock {\em Congress: A political-economic history of roll call voting}.
\newblock Oxford University Press, 1997.

\bibitem{deford_eubank_rodden_2021}
Daryl~R. DeFord, Nicholas Eubank, and Jonathan Rodden.
\newblock Partisan dislocation: A precinct-level measure of representation and
  gerrymandering.
\newblock {\em Political Analysis}, 30(3):403–425, 2021.

\bibitem{Deb_Kelsey_1987}
Rajat Deb and David Kelsey.
\newblock On constructing a generalized ostrogorski paradox: Necessary and
  sufficient conditions.
\newblock {\em Mathematical Social Sciences}, 14(2):161–174, 1987.

\bibitem{Wagner_1984}
Carl Wagner.
\newblock Avoiding anscombe’s paradox.
\newblock {\em Theory and Decision}, 16(3):233–238, 1984.

\bibitem{duchin_2018}
Moon Duchin.
\newblock Gerrymandering metrics: How to measure? what's the baseline?
\newblock {\em arXiv:1801.02064}, 2018.

\bibitem{anes_2020}
American National~Election Studies.
\newblock Anes 2020 time series study full release [dataset and documentation].
\newblock Technical report, ANES, 2021.

\bibitem{lang_xia_2016}
J\'{e}r\^{o}me Lang and Lirong Xia.
\newblock Voting in combinatorial domains.
\newblock {\em Handbook of Computational Social Choice}, page 197–222, 2016.

\bibitem{lacy_2001}
Dean Lacy.
\newblock A theory of nonseparable preferences in survey responses.
\newblock {\em American Journal of Political Science}, 45(2):239, 2001.

\bibitem{riker_brams_1973}
William~H. Riker and Steven~J. Brams.
\newblock The paradox of vote trading.
\newblock {\em American Political Science Review}, 67(4):1235–1247, 1973.

\bibitem{casella_palfrey_2019}
Alessandra Casella and Thomas Palfrey.
\newblock Trading votes for votes. a dynamic theory.
\newblock {\em Econometrica}, 87(2):631–652, Mar 2019.

\bibitem{herra_mace_nunez_2022}
Helios Herrera, Antonin Mac\'{e}, and Mat\'{i}as N\'{u}\~{n}ez.
\newblock Political brinkmanship: Us debt ceiling.
\newblock {\em preprint https://www.heliosherrera.com/Brexit.pdf}, 2022.

\bibitem{arrow_1963}
Kenneth~J. Arrow.
\newblock {\em Social Choice and individual values}.
\newblock Yale University Press, 1963.

\bibitem{Brams_1995}
Steven~J. Brams.
\newblock Approval voting on bills and propositions.
\newblock {\em The Good Society}, 5(2):37–39, Spring 1995.

\bibitem{Brams_Fishburn_1993}
Steven~J. Brams and Peter~C. Fishburn.
\newblock Yes-no voting.
\newblock {\em Social Choice and Welfare}, 10(1), 1993.

\end{thebibliography}

\newpage

\appendix

\section*{Supplementary Information}

%
%
%
%
%
%

\section{Details of the spatial model}

There have been many variations on the spatial model of voting throughout the years. Here, we use one of the simplest models, in which voters are assigned ideal points in the $d$-dimensional hypercube, $[0,1]^d$. Issues are assigned two points, a ``yes'' and a ``no'', and each voter's utility on the issue is determined by the relative distance between her ideal point and the ``yes''/``no'' points. 

\begin{figure}[h]
    \centering
    \includegraphics[width=\textwidth]{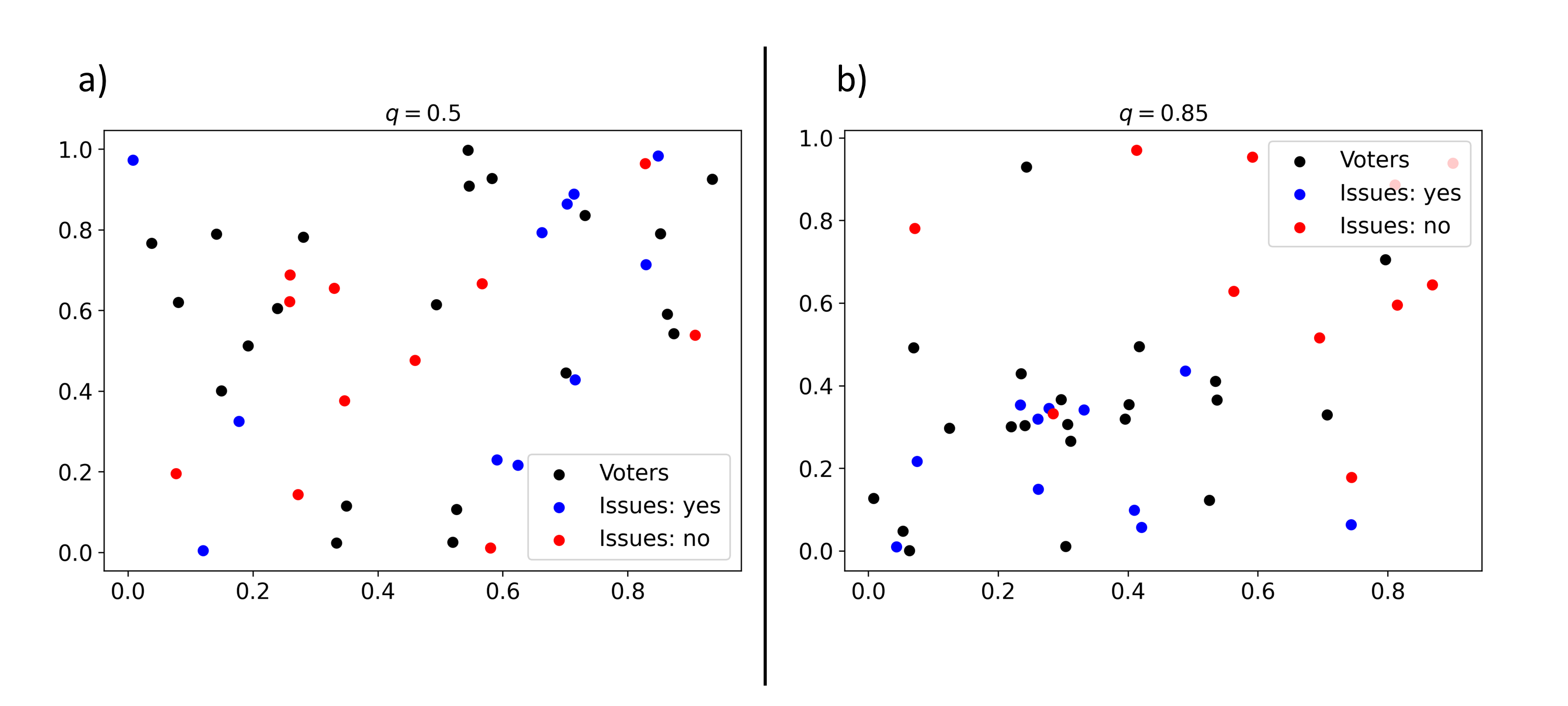}
    \caption{Examples of ideal point distributions for $n=21$, $m=11$. In (a), we see the $q = \frac{1}{2}$ case, where voters and issues are all uniformly spread out in the unit square. In (b), $q=0.85$, so voters and ``yes'' positions are condensed in the bottom left quadrant while ``no'' positions are in the top right. This set of voters and issues will have a large number of positive utilities and only a few negative utilities.}
    \label{fig:spatial_skew}
\end{figure}

In most of the paper, these points are uniformly distributed throughout $[0,1]^n$, as in Figure \ref{fig:spatial_skew}a. When studying the effect of a shifting majority, however, we allow our population to drift to one corner of the hypercube by adjusting a parameter $q$. Each coordinate of a voter's ideal point is uniform between 0 and $\frac{1}{2}$ with probability $q$ and uniform between $\frac{1}{2}$ and 1 with probability $1-q$. Each issue's ``yes'' position is sampled the same way, while the ``no'' positions are reversed, being larger than $\frac{1}{2}$ with probability $q$ and less than $\frac{1}{2}$ with probability $1-q$. Therefore, when $q$ is much greater than $\frac{1}{2}$, the population tends to be in one corner of the hypercube with all the ``yes'' points, while all the ``no'' points are in the opposite corner, and most voters support most issues, as in Figure \ref{fig:spatial_skew}b.

\begin{figure}
    \centering
    \includegraphics[width = \textwidth]{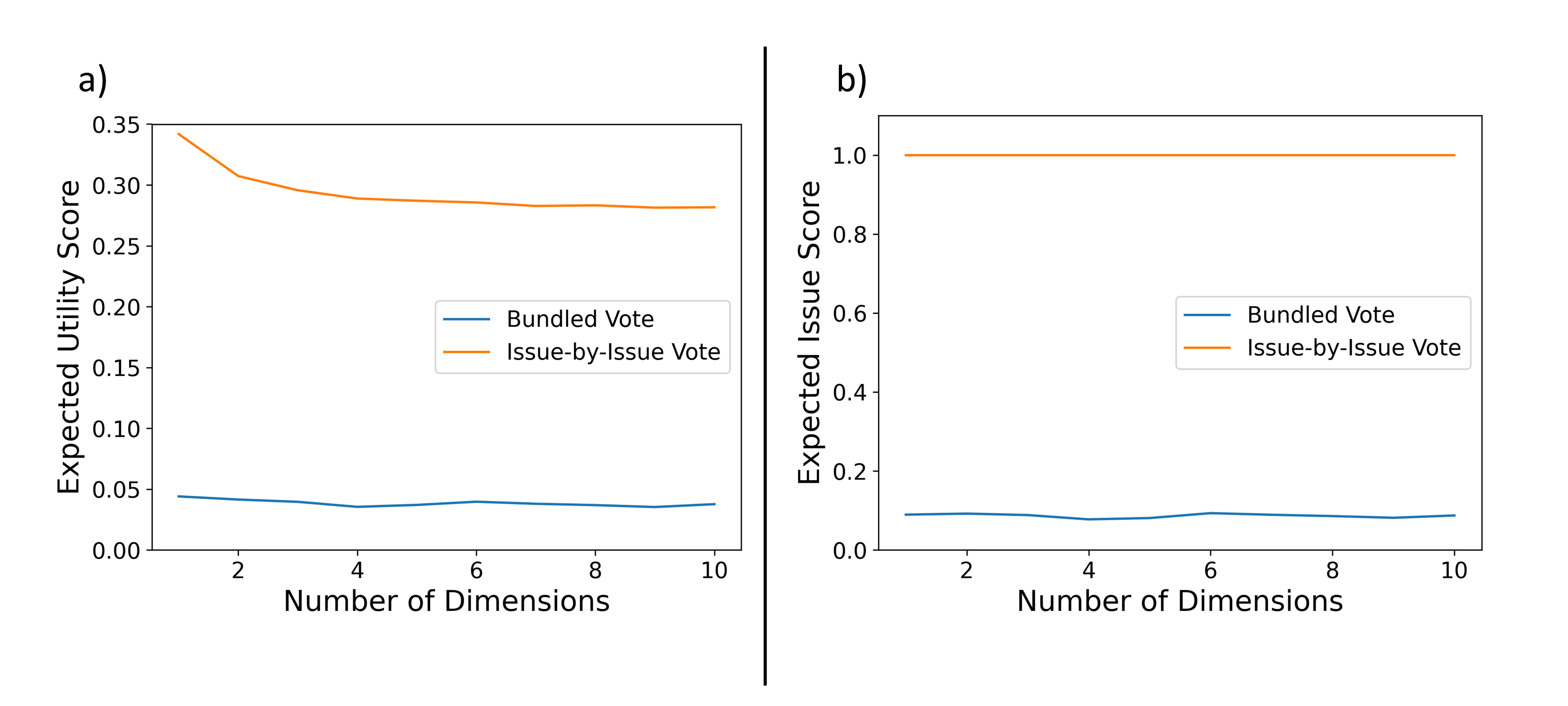}
    \caption{Robustness of bundle value with respect to dimension. As the number of dimensions varies from 1 to 10, the utility (a) and issue (b) scores are both unchanging, suggesting that the choice of dimension does not impact bundle value.}
    \label{fig:vary_dim}
\end{figure}

The use of a spatial model requires choosing the dimension of the underlying space. The natural choice for $d$ is application-specific. A bill deciding whether or not to give extra funding to schools may be fairly one-dimensional if each voter prefers to only fund schools who meet some standardized testing threshold. On the other hand, a group deciding which snacks to get for a party may have many different dimensions, such as sweet vs salty, healthy vs unhealthy, etc. The results in the main paper used $d=2$, and fortunately, this choice has very little impact on the value of the bundle, as can be seen in Figure \ref{fig:vary_dim}.

\section{Derivation of bounds}

Here we derive the upper and lower bounds for the number of number of issues supported in a single bill with $\overline{n}$ supporters, as well as the limits on the number of supporters of the majority and minority supported subbundles.

\subsection{Single Bundle}

Let $n$ be the number of voters considering a bundle of $m$ issues. To avoid ties, assume that $n$ and $m$ are both odd. Suppose the number of voters who vote for the bundle (and therefore support over half the issues) is $\overline{n}$. We would like bounds on the number of issues that are supported, denoted $\overline{m}$. 

\subsubsection{Lower Bound}

In the worst-case scenario, none of the $n-\overline{n}$ voters who did not support the bundle support any of the issues and each of the $\overline{n}$ supporters support the fewest issues possible while still supporting the bundle, $\frac{m+1}{2}$ issues. Therefore, the number of positive utilities in the utility profile can be as low as $\overline{n} \frac{m+1}{2}$.

As far as issues are concerned, each issue can receive $\frac{n-1}{2}$ votes and still be opposed by a majority. Therefore, if the number of positive utilities is less than $m \frac{n-1}{2}$, all issues could have only minority support. In the worst case scenario (for the lower bound), any excess utility past this point is all given to the same issue until that issue has unanimous support to avoid passing more issues than necessary.

\begin{equation}
    \text{Fewest possible total positive utility:  } \overline{n}\frac{m+1}{2}
\end{equation}

\begin{equation}
    \text{Maximum positive utility before any issues pass:  } m \frac{n-1}{2}
\end{equation}

\begin{equation}
    \text{Minimum excess positive utility:  } \overline{n}\frac{m+1}{2} - m \frac{n-1}{2}
\end{equation}

\begin{equation}
    \text{Min num issues passed:  } \ceil*{ \frac{\overline{n}\frac{m+1}{2} - m \frac{n-1}{2}}{\frac{n+1}{2}}
    } = \ceil*{
    \frac{\overline{n}(m+1) - m (n-1)}{n+1}
    }
\end{equation}

Strictly speaking, these final two quantities could be negative (and in fact are for most values of $\overline{n}$), in which case no issues need to be passed.

\subsubsection{Upper Bound}

The computation for the upper bound is similar, except there is no longer any concern about excess positive utility. Instead, ``yes'' votes are all cast for the same issue until it reaches the necessary $\frac{n+1}{2}$ vote threshold to be passed. 

For the upper bound, suppose that the $\overline{n}$ supporters of the bundle support every single issue, and the $n - \overline{n}$ voters that did not support the bundle support $\frac{m-1}{2}$ issues, the maximum possible. Therefore, we have $(n-\overline{n}) \frac{m-1}{2}$ votes from non-supporters. Each issue needs $\frac{n+1}{2} - \overline{n}$ votes from non-bundle-supporters to gain a majority. 

\begin{equation}
    \text{Maximum positive utility:  } \overline{n}m + (n-\overline{n}) \frac{m-1}{2}
\end{equation}

\begin{equation}
    \text{Maximum non-supporter positive utility:  } (n-\overline{n}) \frac{m-1}{2}
\end{equation}

\begin{equation}
    \text{Minimum non-supporter votes needed to pass an issue:  } \frac{n+1}{2}-\overline{n}
\end{equation}

\begin{equation}
    \text{Max num issues passed:  } \floor*{
    \frac{(n-\overline{n})\frac{m-1}{2}}{\frac{n+1}{2} - \overline{n}}
    } = \floor*{
    \frac{(n-\overline{n})(m-1)}{n+1-2\overline{n}}
    }
\end{equation}

\subsubsection{Upper-Lower Bound Duality}

As an interesting note, the placement of positive and negative utility to minimize the number of issues passed with $\overline{n}$ bundle supporters is exactly the opposite of the placement when trying to maximize the number of issues passed with $n - \overline{n}$ supporters. This gives us the clean relationship

\begin{equation}
    \text{lower}(\overline{n}) = \text{upper}(n-\overline{n})
\end{equation}

\subsection{Majority/Minority Supported Subbundles}

Suppose we have $n$ voters and $m$ issues, $m_1$ of which have majority support and $m_2 = m-m_1$ of which do not. We assume $n$ and $m$ are odd to avoid ties, but it is unavoidable that either $m_1$ or $m_2$ will be even.

\textbf{Majority supported issue subbundle}

The majority support subbundle has $m_1$ issues, each with majority support. We want a lower bound on the number of supporters. To begin, we count the number of positive utilities in the subbundle: 

\begin{equation}
    \text{Amount of positive utility} \geq m_1 \left( \frac{n+1}{2} \right).
\end{equation}

Depending on if $m_1$ is even or odd, each voter can support just under half the issues before any voters support the subbundle.

\begin{equation}
    \text{positive utility before any supporters of subbundle} \leq n \left( \frac{m-c}{2} \right)
\end{equation}

where $c = \begin{cases}
1 & m \text{ is odd} \\
2 & m \text{ is even}
\end{cases}.$ 

Therefore, the number of excess votes is bounded by

\begin{equation}
    \text{Amount of excess positive utility} \geq m_1 \left( \frac{n+1}{2} \right) - n \left( \frac{m-c}{2} \right).
\end{equation}

By dividing by the amount of positive utility each voter can take before supporting every issue, we get our final bound:

\begin{equation}
    \text{Number of supporters} \geq \ceil*{
    \frac{m_1 \left( \frac{n+1}{2} \right) - n \left( \frac{m-c}{2} \right)}{\frac{m+c}{2}}
    } = \ceil*{
    \frac{m_1(n+1)-n(m-c)}{m+c}
    }
\end{equation}

\textbf{Minority supported issue subbundle}

This subbundle is similar. Because each issue is supported by a minority, we have the following bound on the number of positive utilities:

\begin{equation}
    \text{Amount of positive utility} \leq m_2\frac{n-1}{2}.
\end{equation}

We also have a lower bound on how much positive utility a supporter needs:

\begin{equation}
    \text{Amount of positive utility per supporter} \geq \frac{m_2+c}{2}
\end{equation}

where $c$ is as above. Dividing these two, we get

\begin{equation}
    \text{Number of supporters} \leq \floor*{
    \frac{m_2 \frac{n-1}{2}}{\frac{m_2+c}{2}}
    } = \floor*{
    \frac{m_2(n-1)}{m_2+c}
    }
\end{equation}

\section{Derivation of bundle value}

In the basic IID model, utility is $+1$ with probability $\frac{1}{2}$ and $-1$ with probability $\frac{1}{2}$, for an expected value of zero. In fact, the expected sum of all the utilities in a utility profile is zero. Despite this, the expected utility score is not zero. To see this, imagine that instead of voting for a bundle, we accept it if it has positive net utility and reject it if it has negative net utility. Both decisions result in positive value, so the expected value from this decision process is positive. While bundled voting may not be quite as effective as this, it still functions the same way, and manages to get a positive expected score by rejecting negative-utility bundles and accepting those with positive utility.

\begin{figure}
    \centering
    \includegraphics[width = \textwidth]{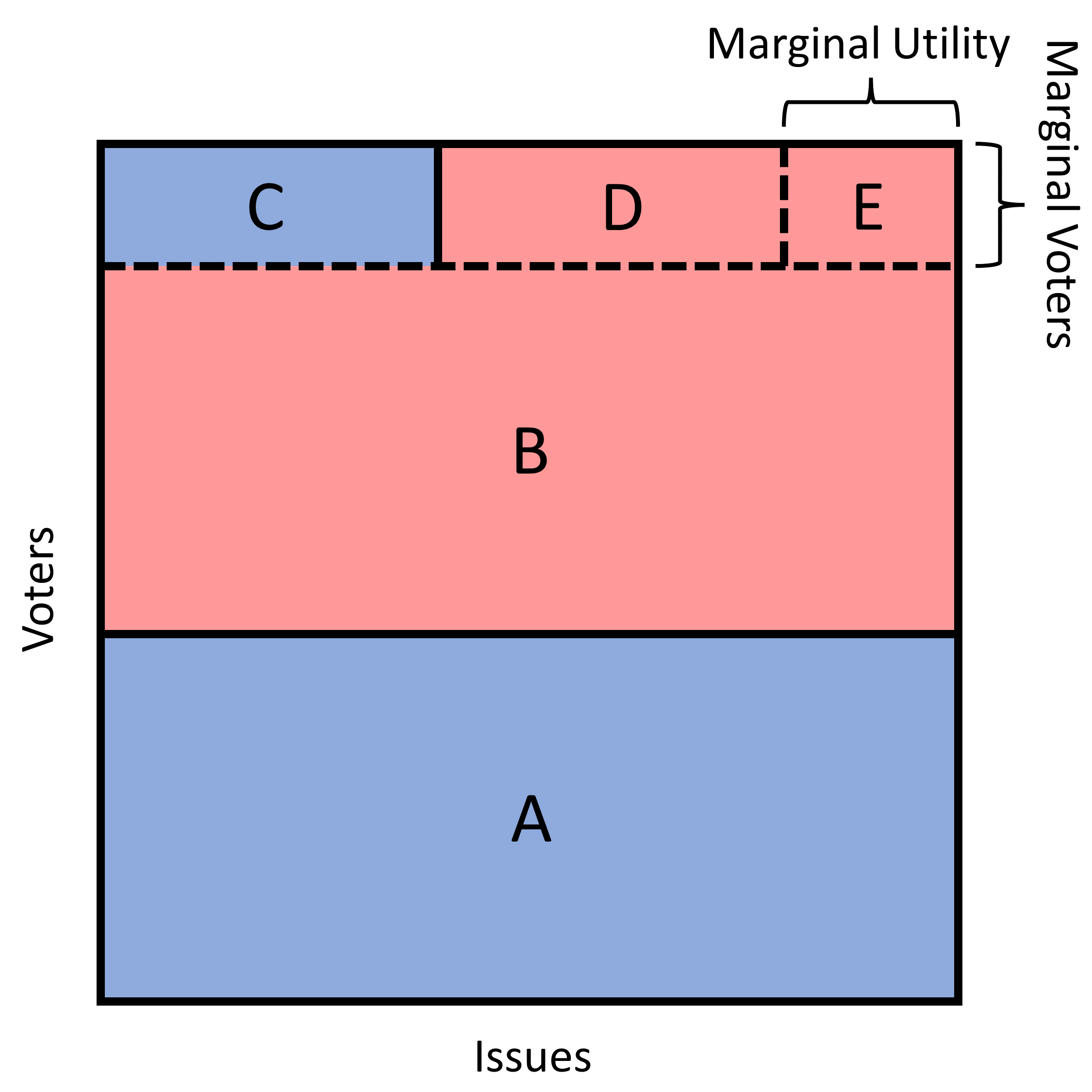}
    \caption{Visual representation of marginal voters and marginal utility. Color represents net positive or negative utility. The voters in A and B have equal size and opposite utility, so they cancel out. The marginal voters are those in blocks C-E. Each marginal voter has some positive and negative utility. The utilities in blocks C and D are equal size and opposite for net zero utility. What remains is block E, the marginal utility of the marginal voters, and this is what provides the positive utility score to the bundled vote. }
    \label{fig:marginals}
\end{figure}

Approximately half of voters will support the bundle and half will reject it. However, by random chance and because $n$ is odd, the number of supporters will be slightly different than the number of detractors. Consider the difference between the number of supports and detractors. If these marginal voters are supporters, the bill passes and they provide positive value. If they are detractors, the bill fails and they \textbf{also} provide positive value (by rejecting a bundle with negative utility). Either way, the non-marginal voters cancel out in expectation (Figure \ref{fig:marginals} blocks A and B) and the marginal voters provide positive value to the bundle.

We can estimate the number of marginal voters by first observing that the number of supporters is binomially distributed, where there are $n$ trials and the probability of success is $\frac{1}{2}$. For large $n$, the binomial distribution $B(n,\frac{1}{2})$ can be approximated by a normal distribution, so the number of supporters is proportional to $\mathcal{N}(n\frac{1}{2}, n\frac{1}{2}\frac{1}{2}) = \mathcal{N}(\frac{n}{2}, \frac{n}{4})$. The normal distribution has expected absolute deviation from the mean of $\sqrt{\frac{2}{\pi}}\sigma$, so the number of expected supporters varies from the mean by $\sqrt{\frac{2}{\pi}}\sqrt{\frac{n}{4}}$. Doubling this gives the expected difference between supporters and detractors, aka the number of marginal voters:

\begin{equation}
    E(\text{Number of marginal voters}) = \sqrt{\frac{2n}{\pi}}.
\end{equation}

We know that each marginal voter votes the same way, and we can use the exact same trick to determine how much value each of these voters adds to the bundle. The number of positive utilities is binomially distributed ($B(m,\frac{1}{2})$) which can be approximated by a normal distribution ($\mathcal{N}(\frac{m}{2}, \frac{m}{4})$) which has an expected absolute deviation from the mean of $\sqrt{\frac{2}{\pi}}\sqrt{\frac{m}{4}}$ which we double to get the marginal utility of voter:

\begin{equation}
    E(\text{Marginal utility of each voter}) = \sqrt{\frac{2m}{\pi}}.
\end{equation}

The rest of the utility for each voter cancels out (Figure \ref{fig:marginals} blocks C and D), so in the end the final utility of the bundle is just the number of marginal voters times the marginal utility per voter:

\begin{equation}
    E(\text{Expected utility of bundle}) = \frac{2}{\pi} \sqrt{mn}.
\end{equation}

The utility score of a bundle is normalized by the maximum possible utility, so the expected utility score is the expected utility divided by $nm$:

\begin{equation}
    E(\text{Expected utility score of bundle}) = \frac{2}{\pi \sqrt{mn}}.
\end{equation}

\section{ANES Data details}

The 11 final issues from the survey data (with their ANES variable name in parentheses) were:

\begin{itemize}
    \item Limits on foreign imports (V202231x)
    \item Change in immigration levels (V202232)
    \item Preferential hiring for blacks (V202252x)
    \item Level of government regulation (V202256)
    \item Government efforts to reduce income inequality (V202259x)
    \item Higher taxes on millionaires (V202325)
    \item Regulation of greenhouse gases (V202336x)
    \item Ban on assault-style weapons (V202344x)
    \item Free trade agreements (V202361x)
    \item Universal basic income of 12k/year (V202376x)
    \item Government spending on healthcare (V202380x)
\end{itemize}

For completeness, the four issues that we considered but discarded were:

\begin{itemize}
    \item Limits on campaign spending (V202225)
    \item Vaccine requirements in schools (V202331x)
    \item Background checks for guns (V202341x)
    \item Government action on opioid addiction (V202350x)
\end{itemize}

\begin{figure}
    \centering
    \includegraphics[width=\textwidth]{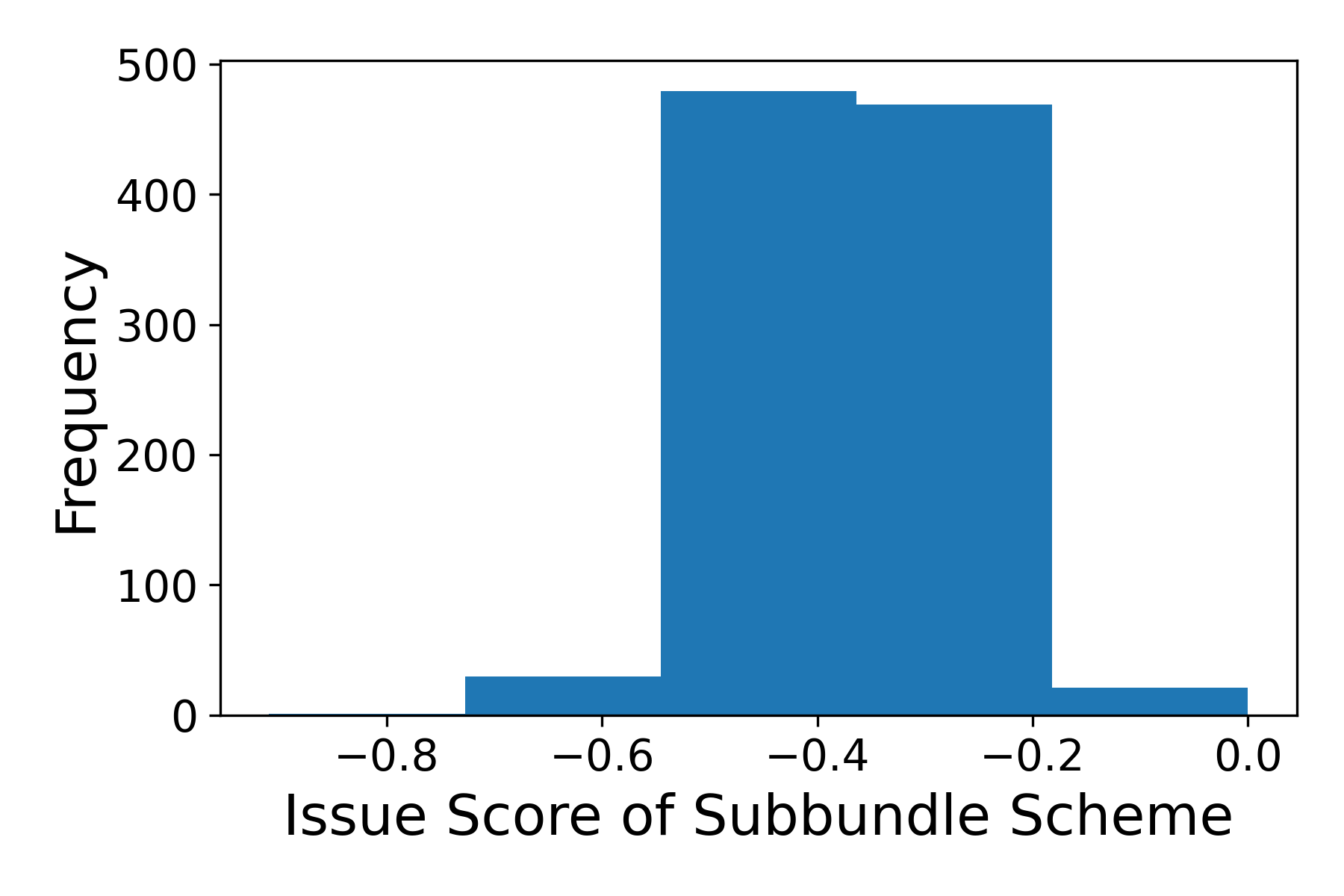}
    \caption{For each sampled group of 31 voters, the distribution of minimum possible subbundle scores across all schemes. The vast majority have issue scores of $\frac{-5}{11}$ or $\frac{-3}{11}$, including the example shown in Figure 5 of the main paper.}
    \label{fig:anes_distribution}
\end{figure}

\begin{figure}
    \centering
    \includegraphics[width = \textwidth]{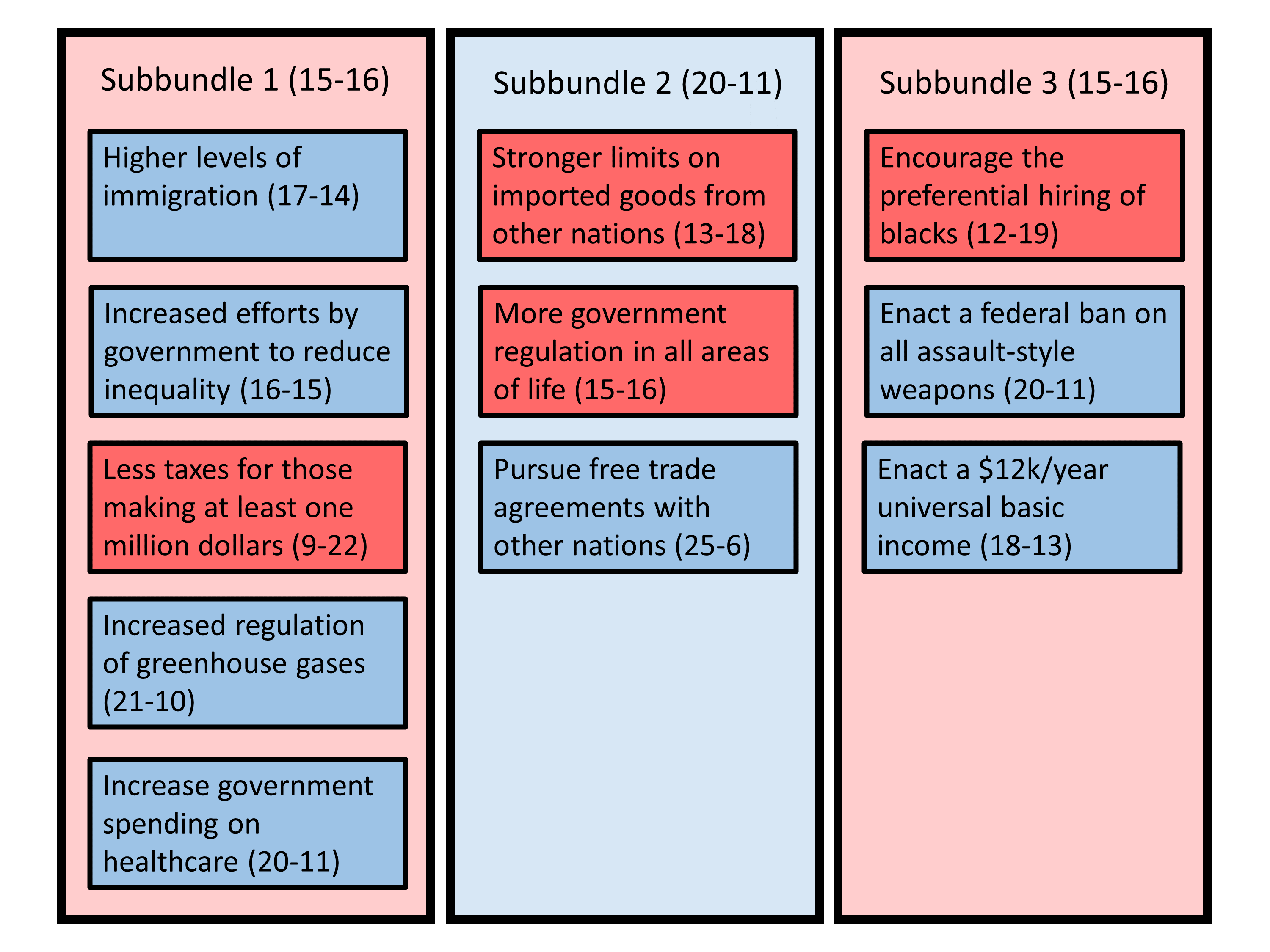}
    \caption{A representative subbundling scheme with an issue score of $\frac{-5}{11}$. Each bundle has one majority-preferred outcome with the rest being beneficial for the NVAAS.}
    \label{fig:representative_gerry}
\end{figure}

The subbundling scheme shown in Figure 5 of the main paper was the most extreme case found. All other subbundle schemes had a lesser degree of manipulation. Figure \ref{fig:anes_distribution} shows the distribution of minimum issue score for each sample of 31 voters. A more representative example of a manipulated subbundle scheme is shown in Figure \ref{fig:representative_gerry}, in which each subbundle has an issue with majority approval.

\end{document}